\documentclass[prb,twocolumn,showpacs,superscriptaddrestitleatfix,10pt]{revtex4-1}

\usepackage{amsmath,amssymb,amsfonts,stmaryrd,wasysym,graphicx,multirow,color,textcomp,subfigure}
\usepackage{url}
\usepackage[colorlinks=true,citecolor=blue,urlcolor=blue]{hyperref}
\hypersetup{colorlinks=true, urlcolor=blue, citecolor=cyan, pdfborder={0 0 0}}

\newcommand{\bea}{\begin{eqnarray}}
\newcommand{\eea}{\end{eqnarray}}
\newcommand{\be}{\begin{eqnarray}}
\newcommand{\ee}{\end{eqnarray}}

\newcommand{\nn}{\nonumber}

\allowdisplaybreaks

\begin{document}

\title{Pressure Induced Topological Superconductivity in the \\ Spin-Orbit  Mott Insulator GaTa$_4$Se$_8$
}

\author{Moon Jip Park}
\thanks{These authors equally contributed.}
\affiliation{Department of Physics, KAIST, Daejeon 34141, Republic of Korea}

\author{GiBaik Sim}
\thanks{These authors equally contributed.}
\affiliation{Department of Physics, KAIST, Daejeon 34141, Republic of Korea}

\author{Min Yong Jeong}
\thanks{These authors equally contributed.}
\affiliation{Department of Physics, KAIST, Daejeon 34141, Republic of Korea}

\author{Archana Mishra}
\affiliation{Department of Physics, KAIST, Daejeon 34141, Republic of Korea}

\author{Myung Joon Han}
\email{mj.han@kaist.ac.kr}
\affiliation{Department of Physics, KAIST, Daejeon 34141, Republic of Korea}

\author{SungBin Lee}
\email{sungbin@kaist.ac.kr}
\affiliation{Department of Physics, KAIST, Daejeon 34141, Republic of Korea}

\date{\today}
\begin{abstract}
{Lacunar spinel GaTa$_4$Se$_8$ is a unique example of spin-orbit coupled Mott insulator described by molecular  $j_{\text{eff}}\!=\!3/2$ states. It becomes superconducting at T$_c$=5.8K under pressure without doping. In this work, we show, this pressure-induced superconductivity is a realization of  a new type topological phase characterized by spin-2 Cooper pairs. Starting from first-principles density functional calculations and random phase approximation, we construct the microscopic model and perform the detailed analysis. Applying pressure is found to trigger the virtual interband tunneling processes assisted by strong Hund coupling, thereby stabilizing a particular $d$-wave quintet channel. Furthermore, we show that its Bogoliubov quasiparticles and their surface states exhibit novel topological nature. To verify our theory, we propose unique experimental signatures that can be measured by Josephson junction transport and scanning tunneling microscope. Our findings open up new directions searching for exotic superconductivity in spin-orbit coupled materials. } 
\end{abstract}

\maketitle
\section*{Introduction}

The confluence of spin-orbit coupling (SOC) and strong electron correlation provides a new paradigm of solid-state quantum phenomena \cite{Sipos2008,PhysRevB.90.041112,Orenstein468,PhysRevLett.112.176402,Kim1329,PhysRevLett.101.076402}. In particular, the new type of superconductivity that are expected to arise in spin-orbit coupled Mott insulators has drawn great attentions. The representative candidate materials are transition metal dicalchogenides TaS$_2$\cite{Sipos2008} and Sr$_2$IrO$_4$\cite{PhysRevX.5.041018,PhysRevLett.101.076402,Gao2015,PhysRevLett.106.136402,Kim1329,PhysRevLett.113.177003,PhysRevLett.110.027002}. Despite of the promising examples, the microscopic superconducting mechanism itself as well as its pairing symmetry remain elusive. The key step forward is to have a concrete material platform for which the unambiguous theoretical description can be provided and tested. In addition, reliable prediction of pairing symmetry and the detailed suggestions for its experimental verification are demanded.

Lacunar spinel compounds, GaM$_4$X$_8$ (M=transition metals; X=chalcogens), are a fascinating class of materials for the demonstration of rich correlated electronic structure and potential applications in technologies \cite{ruff_multiferroicity_2015,reschke_optical_2017,ruff_polar_2017,kezsmarki_ne-type_2015,fujima_thermodynamically_2017,muller2006magnetic,Dorolti2010,PhysRevLett.113.137602,PhysRevB.92.121112,PhysRevB.76.214106}. Among the known lacunar spinels, GaTa$_4$Se$_8$ is a Mott insulator with a charge gap of 0.1$-$0.3eV\cite{pocha_crystal_2005,Guiot2013,guiot_control_2011}. Its widely tunable conductivity is expected to be useful for nonvolatile memory devices \cite{vaju_electric-pulse-driven_2008,guiot_control_2011,dubost_resistive_2013,cario2010electric}. More recent first-principles calculation points out that the SOC of Ta ion induces a novel electronic band structure described by molecular state with so-called $j_{\rm eff}$=3/2 nature \cite{kim_spin-orbital_2014}. 
Subsequently, resonant inelastic x-ray scattering (RIXS) experiment has directly verified this $j_{\rm eff}$=3/2 electronic structure \cite{jeong_direct_2017}, establishing GaTa$_4$Se$_8$ as a notable example of spin-orbit coupled Mott insulator where both electron correlation and SOC play the crucial role.

Strikingly, applying pressure induces the phase transition from a spin-orbit coupled Mott insulator to a metal and eventually to a superconductor\cite{abd-elmeguid_transition_2004,pocha_crystal_2005,ta_phuoc_optical_2013,camjayi_first-order_2014}. 
The characteristics of this superconductivity are quite intriguing in many regards. First, GaTa$_4$Se$_8$ does not show any long-range magnetic order down to low temperature\cite{abd-elmeguid_transition_2004,pocha_crystal_2005}. 
Second, there is no experimental signature for structural transition as a function of pressure and no drastic phonon mode change. Third, nevertheless, the anomalies in specific heat as well as magnetic susceptibility are repeatedly identified at  around 50K which is an order of magnitude higher than superconducting T$_c$
\cite{pocha_crystal_2005,jakob_structural_2007,yaich_nouveaux_1984,kawamoto_frustrated_2016,waki_spin-singlet_2010}. Most importantly,
it is also noted that superconductivity is only observed in the case of M=Nb and Ta; namely, only when the low energy band structure is of $j_{\rm eff}$=3/2 character \cite{kim_spin-orbital_2014}. This observation heavily prompts a speculation that $j_{\rm eff}$=3/2 nature of the electronic band structure pervades the origin of the superconductivity, being different from the conventional BCS type. However, there has been no firm investigation on its character both theoretically and experimentally.

\begin{figure*}
	\includegraphics[scale=0.8]{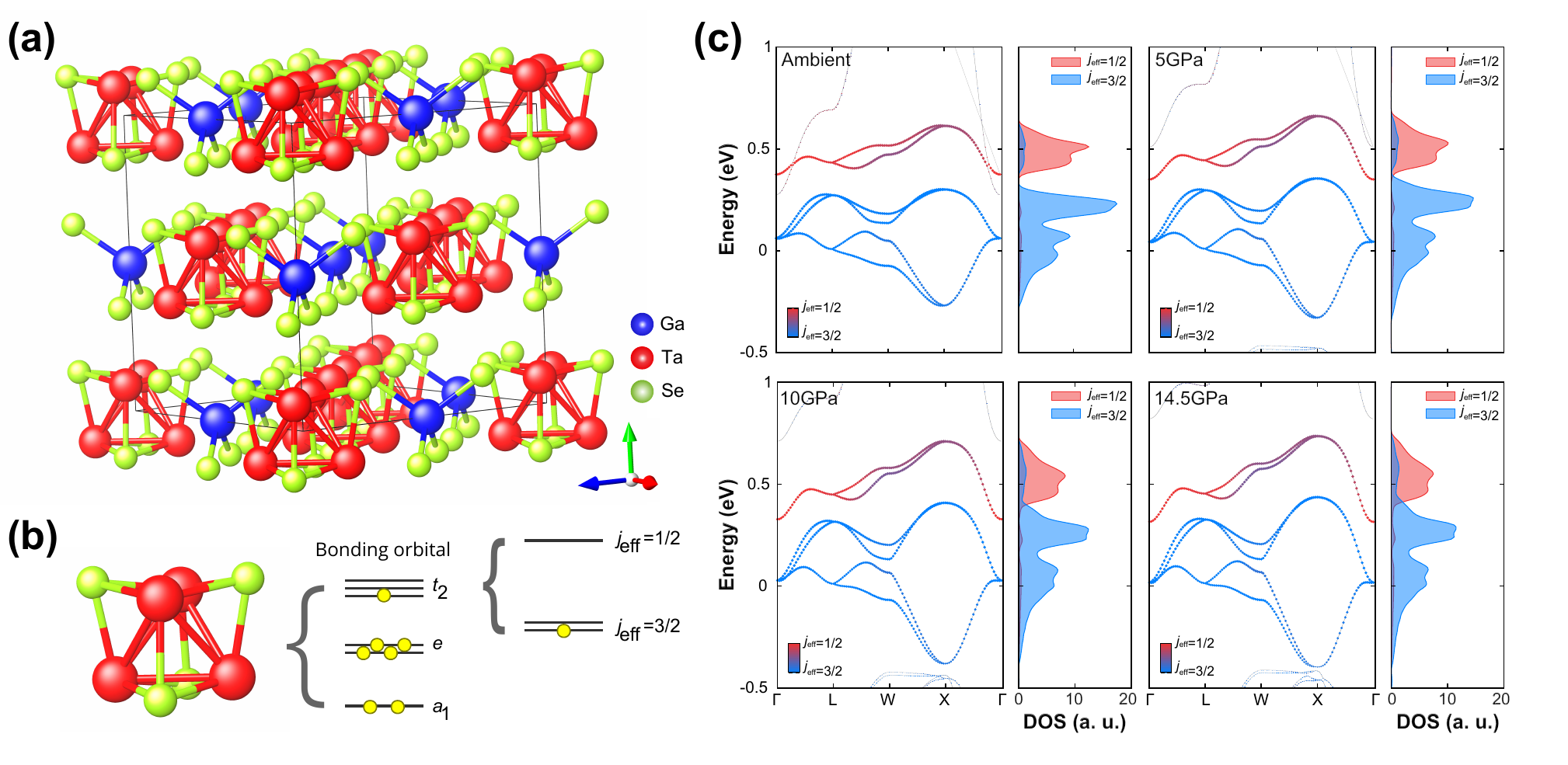}
	\caption{(a) Crystal structure of GaTa$_4$Se$_8$. GaSe$_4$ and Ta$_4$Se$_4$ clusters consist of NaCl-like structure. (b) Schematic electronic structure near Fermi level. Molecular bonding orbitals $e$ and $a_1$ are fully-filled and one electron is in $t_2$ orbitals. By SOC, $t_2$ orbitals split into $j_{\rm{eff}}$=1/2 doublet and $j_{\rm{eff}}$=3/2 quartet with one electron in $j_{\rm{eff}}$=3/2 bands. (c) Fat bands and PDOS of GaTa$_4$Se$_8$ for various pressures (ambient, 5GPa, 10GPa and 14.5GPa). Even under high pressure, $j_{\rm{eff}}$=3/2 bands are well separated from other bands.}
	\label{intro_dft}
\end{figure*}


In this paper, we show that the superconductivity in GaTa$_4$Se$_8$ is attributed to the new type of electronic pairing. Due to the intriguing interplay of multi-band $j_{\rm eff}=3/2$ character and inter-band correlation, novel $d$-wave quintet superconductivity with spin-2 Cooper pairs is stabilized. Such high angular momentum Cooper pair state has been also referred to as the quintet pairing states\cite{PhysRevB.100.224505,PhysRevB.96.214514,PhysRevLett.120.057002,PhysRevLett.118.127001,PhysRevB.99.054505,PhysRevX.8.011029,PhysRevLett.116.177001}. Utilizing both density functional theory and random phase approximation (RPA), we first show that the system well retains the characteristic of $j_{\rm eff}$=3/2 under high pressure. Our first-principles calculations also show how intra-, inter-orbital electron interactions and Hund coupling change by pressure. Starting from the constructed many-body Hamiltonian, we analytically show that applying pressure activates many-body inter-band tunnelings and opens attractive quintet pairing channels assisted by strong Hund coupling. Among the possible quintet pairings, it turns out the system favors a particular $d$-wave superconductivity with $t_{2g}$ symmetries. This novel superconductivity is characterized by nodal lines of Bogoliubov quasiparticles and by topologically protected Majorana modes at the surface. Thereby, our work theoretically establishes GaTa$_4$Se$_8$ as a strong candidate of topological $d$-wave superconductor. In order to facilitate its confirmation, we also propose the concrete experimental setups and the signatures to be identified in Josephson junction transport and scanning tunneling microscopy (STM).

\section*{Results}

\subsection*{Electronic Structure and Many-Body Hamiltonian}

GaTa$_4$Se$_8$ consists of GaSe$_4$ and Ta$_4$Se$_4$ clusters arranged in NaCl structure (see Fig.~\ref{intro_dft}(a)) \cite{pocha_crystal_2005} that belong to the space group F$\bar{4}3$m, which forms non-centrosymmetric structure. Due to the short intra-cluster bondings, its electronic band structure is well understood by molecular orbital states, and the states near Fermi level are dominated by triply degenerate molecular $t_2$ orbitals denoted by ($D_{xy},~D_{yz},~D_{zx}$) \cite{pocha_crystal_2005}. Just as the atomic $t_{\rm{2g}}$ orbitals, molecular $t_2$ can also be represented by effective angular momentum $l_{\rm{eff}}=1=-L_{(1)}$ where $L_{(1)}$ is the angular momentum operator with orbital quantum number $l=1$\cite{chen2011spin}.
The spin-orbit interaction, $H_{\rm SOC}\!=\!-\lambda l \cdot S$, gives rise to the molecular quartet $j_{\rm eff}=3/2$ and the doublet $j_{\rm eff}=1/2$. In particular, molecular quartet $j_{\rm eff}=3/2$ in the basis of $| j, j^z>$ is being represented as
$
|3/2,\pm 3/2>=\mp\frac{ 1}{\sqrt{2}}(|D_{yz,\uparrow\downarrow}\rangle \!\pm\! i|D_{zx,\uparrow\downarrow}\rangle)$ and 
$|3/2,\pm 1/2>=\sqrt{\frac{2}{3}}(|D_{xy,\uparrow\downarrow}\rangle\mp\frac{|D_{yz,\downarrow\uparrow}\rangle \!\pm\! i|D_{zx,\downarrow\uparrow}\rangle}{2})$ where $\uparrow,\downarrow$ refer to spin directions \cite{kim_spin-orbital_2014}.

The calculated band dispersions and the projected density of states (PDOS) are shown in Fig.~\ref{intro_dft}(c) as a function of pressure (for more details, see Supplementary Information \ref{Appendix:DFT} and Ref. \onlinecite{pocha_crystal_2005} for the crystal structure data under the pressures). Note that, not only at the ambient pressure but at the high pressure up to 14.5 GPa, $j_{\rm{eff}}$ = 3/2 band characters are well maintained and still dominating the near Fermi energy region. It justifies our low energy model containing $j_{\rm{eff}}$ = 3/2 states.

\begin{figure*}
	\centering\includegraphics[width=0.8\textwidth]{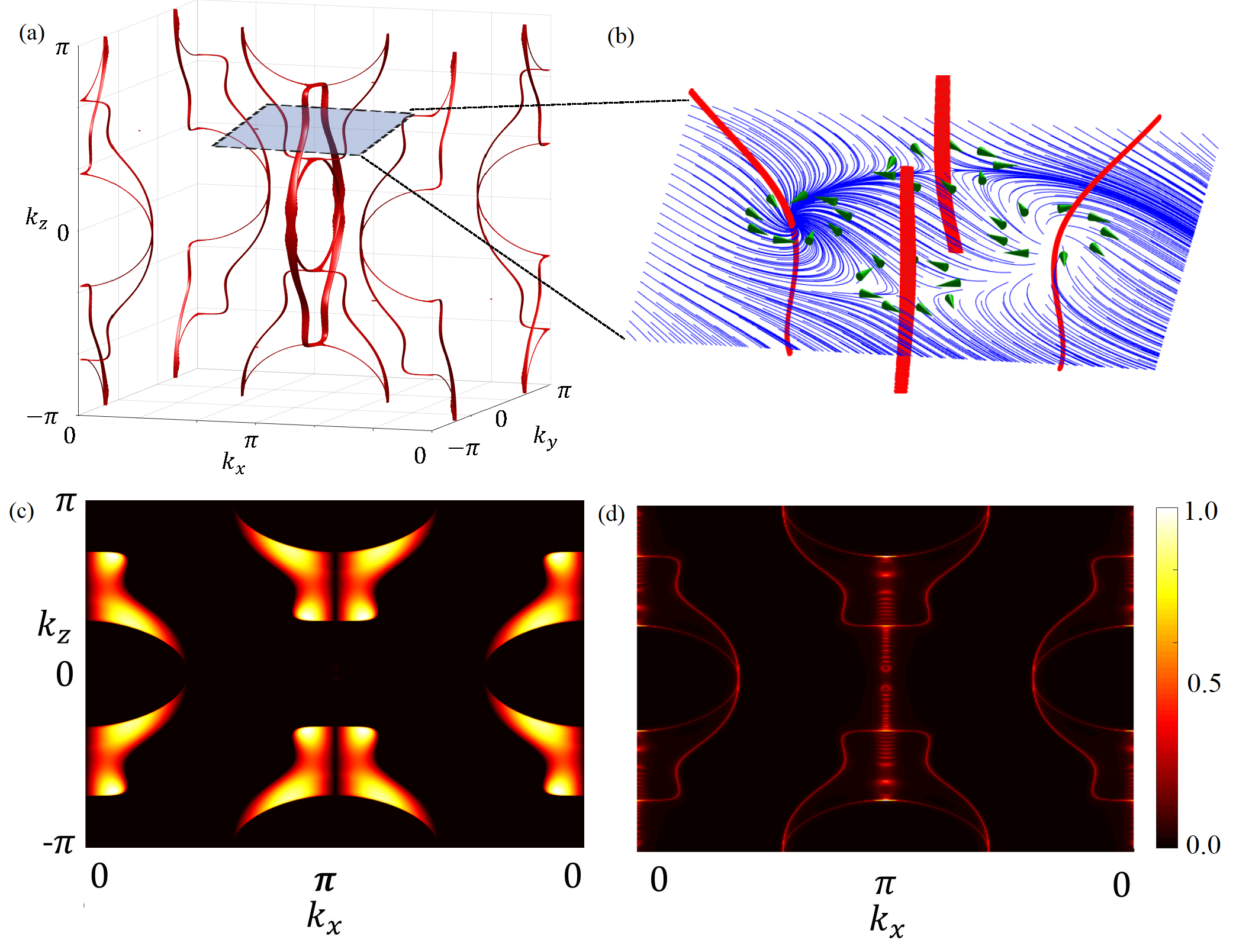}
	\caption{ (a) The BdG gap structure of $d_{xy}$ pairing. We find two-fold degenerate nodal line gap structure shown as the red line. The red columns indicate the nodal lines. (b) The non-trivial winding of phase $\Phi(\mathbf{k})$ around the nodal lines. The blue streamline represents the winding of the phase $\Phi(\mathbf{k})$. We find that the non-trivial winding number protects the nodal line. (c)-(d) The normalized zero-energy spectral density of the surface(c) and bulk(d). Open surface possesses the topological Majorana flat band covering the interior of the nodal line. The open boundary condition is taken along $[0 1 0]$-direction. 
	}\label{fig:SC1} 
\end{figure*}

\begin{figure*}
	\centering\includegraphics[width=0.8\textwidth]{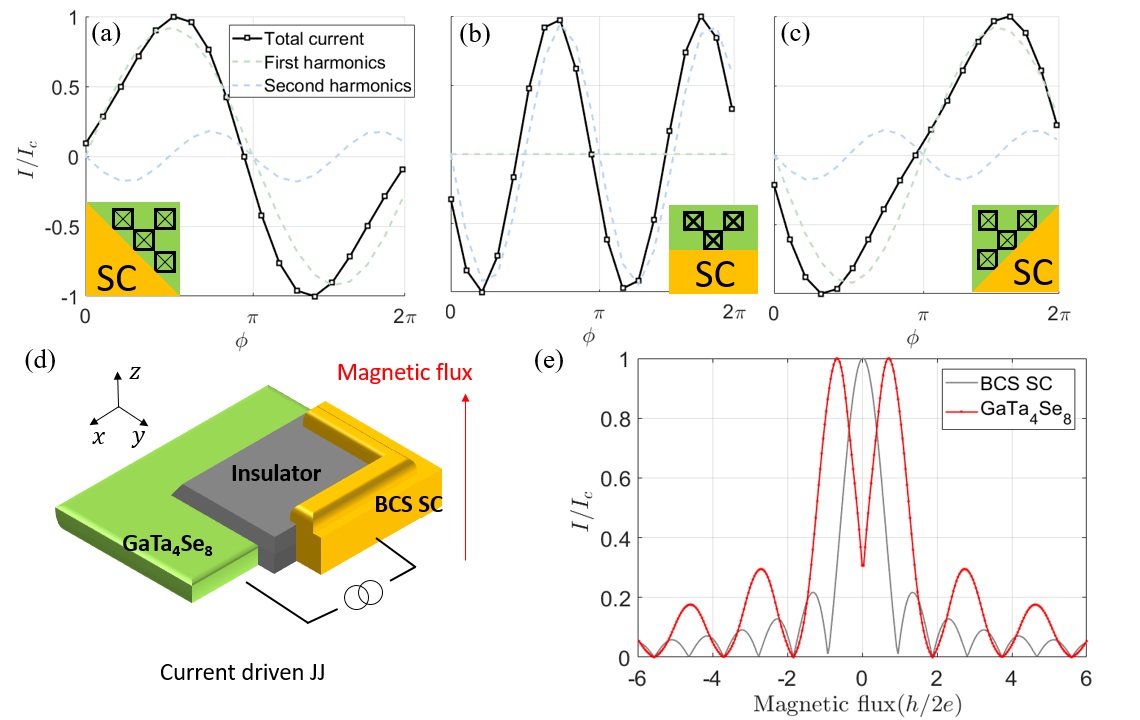}
	\caption{ (a-c) The CPR of the planar Josephson junction at different orientations. Due to the $d_{\rm xy}$ pairing symmetry, the first harmonics of the CPR is inverted under $90^\circ$ rotation.  (d) Schematic figure of the Josephson corner junction. (e) The unconventional Fraunhofer pattern of the lacunar spinel. Unlike the conventional Fraunhofer pattern, the location of the peaks and the dips are reversed.
	} 
	\label{Fig:exp}
\end{figure*}

In order to take into account electronic correlations, we construct many-body Hamiltonian including intra-orbital ($U>0$), inter-orbital interaction ($U'>0$) and Hund coupling ($J_{\rm{H}}>0$):
\begin{align}
&H_I= U \sum_{u}n_{u\uparrow}n_{u\downarrow}+U'\sum_{u,v<u}n_{u\sigma}n_{v\sigma'} \nonumber \\
&+\! \frac{J_{\rm{H}}}{2}\! \!  \sum_{u\neq v ,\sigma,\sigma'}d^{\dagger}_{u\sigma}d^{\dagger}_{v\sigma'}d_{u\sigma'}d_{v\sigma}
\! +\! \frac{J_{\rm{H}}}{2} \! \! \sum_{u\neq v ,\sigma\neq\sigma'}d^{\dagger}_{u\sigma}d^{\dagger}_{u\sigma'}d_{v\sigma'}d_{v\sigma} 
\label{eq:H_I}
\end{align}
where $d_{iu\sigma} ~(d^\dagger_{iu\sigma})$ is the annihilation (creation) operator of electrons with orbital $u\in (D_{xy},D_{yz},D_{zx})$ and spin $\sigma$. 
The third and fourth terms are the Hund exchange and Hund's pair hopping interaction, respectively.

\subsection*{$d$-wave Quintet Pairing}

Projecting this many-body interactions onto $j_{\rm eff}=3/2$ basis (namely, $\Psi=(|3/2 \rangle,|1/2 \rangle,|-1/2 \rangle,|-3/2 \rangle)$) and using Fierz transformation, Eq.\eqref{eq:H_I} is exactly decoupled into the singlet and the five distinct quintet pairing channels as follow:
\begin{align}
H_{I,J_{\rm eff}=3/2}&=g_0(\Psi^\dagger T^\dagger \Psi^*)(\Psi^T T\Psi) \nonumber \\
&+
g_1\sum_{a=1}^3(\Psi^\dagger(T\gamma_a)^\dagger\Psi^*)(\Psi^T T\gamma_a\Psi) \nonumber \\
&+ 
g_2\sum_{a=4}^5(\Psi^\dagger(T\gamma_a)^\dagger\Psi^*)(\Psi^T T\gamma_a\Psi).
\label{totintham1}
\end{align}
Here, $\gamma_a$ is the four-dimensional gamma matrices representing quintet-spin operator and $T$ is the unitary component of the time-reversal operator (for more details, see Supplementary Information \ref{appendix:intraband}). $g_{1}$ and $g_{2}$ represent the quintet pairing strength with $t_{2g}$ and $e_{g}$ symmetry, respectively, while $g_0$ is the singlet pairing strength. Assuming that $j_{\rm eff}=3/2$ states are well separated from $1/2$ bands, one can have the exact expression of each coupling constant: $g_0=({2U+U'+3J_{\rm{H}}})/{24}$, $g_1=(3U'-J_{\rm{H}})/24$ and $g_2=(U+2U'-3J_{\rm{H}})/{24}$\cite{PhysRevB.99.024516}.

It is remarkable that no matter how large is the intra-orbital interaction $U$, $g_1$ coupling can be attractive and therefore induce superconducting instability if the Hund coupling is comparable to inter-orbital interaction, $J_H>3U'$. In contrast, the singlet pairing channel cannot be attractive ($g_0<0$) since the Hund’s coupling and the Hubbard interactions are both positive. This result is irrespective of the interaction parameters, which single out the possibility of the trivial singlet superconductivity.

Importantly, this $t_{2g}$ symmetry $d$-wave pairing is robust even when the inter-band mixings between $j_{\rm{eff}}=3/2$ and $1/2$ are considered.
Fig.~\ref{intro_dft}(c) shows the band separation between $j_{\rm{eff}}=3/2$ and $1/2$ gradually decreases as the pressure increases the bandwidth. At high enough pressure, the sizable many-body interband tunneling is expected. In this regime, $j_{\rm eff}=1/2$ can make an additional contribution to the effective pairing interaction, $g$, through the virtual tunneling process. This effect can formally be calculated using many-body Schrieffer-Wolff transformation\cite{BRAVYI20112793}. Interestingly, we find, the leading order contribution of the interband tunneling is always attractive pairing interactions irrespective of the specific values of $(U,U',J_H)$ (See Supplementary Information \ref{appendix:interband} for the estimation of the interaction parameters derived using the RPA calculation.) As a result, the tunneling effect, assisted by strong $J_H$, opens up the attractive superconducting channel characterized by $g_1<0$, and results in  quintet-spin Cooper pairs with $t_{2g}$ $d$-wave symmetry.

\subsection*{Topological Superconductivity}
The intriguing nature of this $d$-wave quintet Cooper pair can be found in its non-trivial spin texture originating from the unique topological property of Bogliubov-de Gennes (BdG) energy spectrum. Among the possible superconducting order parameter configurations with $t_{2g}$ symmetry, we find that the energetically most favorable state, $\langle \psi^T T \vec{\gamma}\psi\rangle=(1,0,0)$-state where $\vec{\gamma}=(\gamma_1, \gamma_2, \gamma_3)$ represents quintet pairing with $t_{2g}$ symmetry, is characterized by gapless nodal lines as shown in Fig.\ref{fig:SC1} (a); see Supplementary Information \ref{appendix:meanfield} for more details of our calculation.
Due to the $j_{\rm eff}=3/2$ orbital character, the nodal lines in Fig.~\ref{fig:SC1}(a) exhibit robust $d_{xy}$ symmetry even in the presence of small inversion symmetry breaking terms. This nodal lines have a topological origin and are protected by the non-trivial winding number.

It is first noted that the particle-hole and the time-reversal symmetry allow us to define the following non-hermitian matrix and its singular value decomposition, $h_0(\mathbf{k})+i T|\Delta|\gamma_1\equiv U_\mathbf{k}^\dagger D_\mathbf{k} V_\mathbf{k}$, where $h_0(\mathbf{k})$ is the normal Hamiltonian. $D_\mathbf{k}$ is now a diagonal matrix containing all the positive energy eigenvalues. Second, we can consider the adiabatic band flattening process by smoothly deforming $D_{\mathbf{k}}$ to $\mathbb{I}_4$ without any gap closing. This procedure defines the new unitary matrix, $q_\mathbf{k}\equiv U_\mathbf{k}^\dagger V_\mathbf{k}=\sum_n e^{i\lambda_{n}(\mathbf{k})}|n(\mathbf{k}) \rangle \langle n(\mathbf{k})|$, and the corresponding phase  $\lambda_{n}(\mathbf{k})$. These phases are well-defined as long as the system is fully gapped. Therefore, one can assign $\mathbb{Z}_2$ topological winding number along a line that encircles the nodal line as follow:
$w=\frac{i}{2\pi}\oint d\mathbf{k}\cdot tr(q_\mathbf{k}^\dagger \nabla_\mathbf{k} q_\mathbf{k})$ according to DIII class in the Altland-Zirnbauer classifications\cite{RevModPhys.88.035005}.

Fig.~\ref{fig:SC1}(b) shows the configuration of the phase $\Phi(\mathbf{k})\equiv\sum_{n}\lambda_n(\mathbf{k})$. Blue streamlines clearly show that the phases have vortex-antivortex configurations where the core of the vortex defines the nodal line. From the explicit calculation of the winding number, we conclude that each nodal line and the vortex configuration are topologically characterized by the non-trivial winding number, $w=\pm1$. These vortical configurations cannot be removed unless the vortex-antivortex pair annihilates each other. Thus, the nodal lines are topologically protected.

\subsection*{Experimental Verifications}
We now suggest the experimental signatures to verify  $d$-wave quintet pairing. First of all, the non-trivial winding number encircling the nodal line manifests itself as the Majorana zero modes on the open surface. Fig.~\ref{fig:SC1}(c) shows the simulated Majorana flat band, which can be directly observed by STM \cite{Wang333} and superconducting tunneling spectroscopy \cite{PhysRevB.95.174505,PhysRevB.93.201105}. The Majorana zero-modes depicted in Fig.~\ref{fig:SC1}(c) exist in every momentum point in the interior of the surface projected nodal line. Thus the Majorana flat band contributes to the zero-energy density of state at the surface.

Another experiment we suggest is Josephson junction transport. Fig.~\ref{Fig:exp}(a)-(c) show the current-phase relation (CPR) for the planar junction of rotating orientations. The CPR can be expressed as a series of sinusoidal harmonics of the phase difference, $\phi$: $I_J(\phi)=\sum_n I_n \sin(n\phi)$ where $I_n$ gives the $2\pi n$ periodic Josephson current component. Due to the $d_{xy}$ pairing symmetry, Josephson coupling gains $\pi$ phase under $90^\circ$ rotation of the junction orientation, and therefore the sign of $I_1$ is inverted as shown in Fig.~\ref{Fig:exp}(a) and (c). In between the two angles ({\it i.e.}, when the junction is formed along the [100]-direction), the first harmonics vanishes, $I_1=0$; see Fig.~\ref{Fig:exp}(b).  The next dominant CPR has $\pi$ periodicity and the resulting Josephson frequency, $4eV/h$, is the twice of the conventional Josephson frequency \cite{PhysRevB.98.104515}. This frequency doubling can be directly observed from the measurement of the Shapiro step in the I-V characteristics.

One can also make use of the pairing symmetry in this material which results in the unconventional magnetic oscillation pattern \cite{PhysRevLett.71.2134,PhysRevLett.74.797,RevModPhys.67.515,PhysRevB.96.064518,Chen2018}. Fig.~\ref{Fig:exp}(d) shows the schematic setup of the Josephson corner junction which is constructed on the corner of the lacunar spinel crystal. Due to the $\pi$-phase difference in CPR with different orientations, Josephson currents at each face destructively interfere with each other. However, because of small inversion symmetry breaking in the system, the critical current does not completely cancel but makes the dips in the Fraunhofer diffraction pattern as shown in Fig.\ref{Fig:exp}(e). As a consequence, we find that the overall locations of the peaks and the dips in the Fraunhofer pattern should be reversed compared to the case of conventional superconductors. This unusual magnetic oscillation pattern can be regarded as the signature of the quintet pairing in GaTa$_4$Se$_8$.

\section*{Discussion}

We discuss the relevance to another lacunar spinel material GaNb$_4$Se$_8$ which shares many similar features with GaTa$_4$Se$_8$. At ambient pressure, GaNb$_4$Se$_8$ is known to have a Mott gap of 0.19 eV \cite{pocha_crystal_2005}, and the previous calculation shows that its low energy band character is also well-identified by $j_{\rm eff}$=3/2 states due to the sizable SOC in Nb atoms \cite{kim_spin-orbital_2014}. The pressure-induced superconductivity is also found with $T_c=2.9$K at $13$GPa\cite{abd-elmeguid_transition_2004}. Such similarity with GaTa$_4$Se$_8$ may indicate GaNb$_4$Se$_8$ as another strong candidate of topological superconductors. Nevertheless, in contrast to GaTa$_4$Se$_8$, GaNb$_4$Se$_8$ has a sizable band overlap between $j_{\rm{eff}}=3/2$ and $j_{\rm{eff}}=1/2$ bands\cite{kim_spin-orbital_2014}, which indicates the stronger inter-band tunneling effect that goes beyond the analysis of Schrieffer-Wolff transformation method. While stronger virtual tunneling effect on the superconductivity is expected, detailed correlation effect may be different under pressure, depending on molecular states occupied with either Nb or Ta.

In addition, Guiot et al. have performed Te doping in GaTa$_4$Se$_8$ by substitution of Se atoms\cite{guiot_control_2011}. The empirical effect of the Te doping is the reduction of the effective bandwidth followed by the increase of the Mott gap. Similarly, we may expect the increase of superconducting critical temperature. This would solidify our prediction that the superconducting pairing mediated by the electron-electron interaction than the phonon coupling. 

Furthermore, the pressure control can be another interesting path to control the superconducting phase transition. Near the superconducting critical point, the transition to the time-reversal broken states is expected (See supplementary information \ref{appendix:meanfield}). The time-reversal broken states is signatured by the Bogoliubov Fermi surfaces, and it can be measured by the anomalous thermal Hall effect similar to that of the $p+ip$ chiral superconductor. In general, the time-reversal broken phase cannot occur in conventional singlet pairing. Therefore, the thermal Hall effect near the superconducting phase transition would be another smoking gun signature of the quintet superconductivity.

In summary, we have suggested a new superconducting pairing mechanism for which spin-orbit entangled multiband nature plays an essential role together with electron correlation. Our theory is developed for and finds its relevance to GaTa$_4$Se$_8$ and other lacunar spinels, where the origin of pressure induced superconductivity has not been understood for a long time. Starting from the realistic band structure and considering the correlation strengths calculated by first-principles DFT calculations, we have developed the detailed microscopic theory. Superconducting gap is found to have $d$-wave symmetry and its gapless nodal lines emerge with the non-trivial topological character. Furthermore, we have proposed concrete experiments that can confirm our theoretical suggestion. The unusual $I$--$V$ characteristics and the magnetic oscillation patterns are expected from Josephson transport and can be regarded as the smoking gun signatures for this quintet paring. STM image can also be compared with our results. Our findings will pave a new way to search for exotic superconductivity in lacunar spinel compounds.

\section*{Methods}
\subsection*{First-principles calculation}
Electronic structures calculations were performed with OPENMX software package based on linear combination of pseudo-atomic-orbital basis\cite{ozaki_variationally_2003} and within local density approximation (LDA) \cite{ceperley_ground_1980,perdew_self-interaction_1981}. The SOC was treated within the fully relativistic j-dependent pseudo-potential scheme \cite{macdonald_relativistic_1979}. We used the $12\times12\times12$ k-grids for momentum-space integration and the experimental crystal structures at different pressures \cite{pocha_crystal_2005}. For the estimation of tight-binding hopping and interaction parameters, we used maximally localized Wannier function (MLWF) method \cite{marzari_maximally_1997,souza_maximally_2001} and constrained RPA (cRPA) technique \cite{aryasetiawan_frequency-dependent_2004,PhysRevB.77.085122,sasioglu_effective_2011} as implmented in ECALJ code \cite{kotani_ecalj_nodate}.

\section{Data availability}
The data that support the findings of this study are available from the corresponding author on reasonable request.

\section{Code availability}
The computer code used for this study is available upon reasonable request.

\section*{Acknowledgement}
M.Y.J. and M.J.H. are grateful to Jungho Kim, Seo Hyoung Chang, and Seung Woo Jang for fruitful discussion. M.Y.J. and M.J.H. were supported by Basic Science Research Program (2018R1A2B2005204) and Creative Materials Discovery Program (2018M3D1A1058754) through the National Research Foundation of Korea (NRF) funded by the Ministry of Science and ICT. M.J.P., G.B.S. and S.B.L. are supported by the KAIST startup, BK21 and National Research Foundation Grant (NRF-2017R1A2B4008097).

\section*{Competing interests}
The authors declare no conflict of interest.

\section*{Author contributions}
S.B.L. and M.J.H. conceived and supervised the research. M.J.P., G.B.S. and M.Y.J performed the calculations in this work. A.M. provided preliminary results regarding the Fierz transformation, and contributed to the discussion of the quintet pairing superconductivity. All authors contributed to writing the manuscript. 

\clearpage
\pagebreak
\newpage

\renewcommand{\thesection}{\arabic{section}}
\setcounter{section}{0}
\renewcommand{\thefigure}{S\arabic{figure}}
\setcounter{figure}{0}
\renewcommand{\theequation}{S\arabic{equation}}
\setcounter{equation}{0}
\begin{widetext}

	\section{\textit{ab initio} parameters and DFT calculation details}\label{Appendix:DFT}

The electronic structure calculations by using OPENMX were carried out with 400 Ry energy cutoff. In order to take into account of pressure effect, we used the experimental lattice parameters measured at  0 (ambient), 5, 10, and 14.5 GPa \cite{pocha_crystal_2005}. The tight-binding hopping parameters and the interaction parameters were estimated in between $t_2$ molecular orbitals. As shown in Fig.~\ref{fig:crpa}(a), MLWF-based tight-binding bands well reproduce the DFT-LDA results. Fig.~\ref{fig:MLWF}(a) visualizes  the calculated MLWFs denoted by $D_{\rm{xy}}$, $D_{\rm{yz}}$, and $D_{\rm{zx}}$ each of which is composed of four atomic $d_{\rm{xy}}$, $d_{\rm{yz}}$, and $d_{\rm{zx}}$ orbitals, respectively. Four major hopping parameters are presented in Table.~\ref{table_tb} where we present the values obtained from ECALJ code \cite{kotani_ecalj_nodate}. This set of hopping parameters were double checked with OPENMX, and the deviations are found to be less than 1 meV.

\begin{table*}[!h]
	\centering
	\begin{tabular}{|c|c|c|c|c|c|}
		\hline
		\bf{Lattice structure} & \bf{Pressure (GPa)} & $\bf{t_1}$\bf{(meV)} & $\bf{t_2}$\bf{(meV)} & $\bf{t_3}$\bf{(meV)}&$\bf{t'}$\bf{(meV)}\\
		\hline
		\multirow{4}{*}{\bf{Experimental}} & $0$ & $-55$ & $28$ & $7$ & $14$\\
		\cline{2-6}
		& $5$ & $-65$ & $31$ & $7$  & $14$\\
		\cline{2-6}
		& $10$ & $-73$ & $35$ & $5$  & $15$\\
		\cline{2-6}
		& $14.5$ & $-78$ & $36$ & $4$  & $15$\\
		\hline
		\bf{Optimized (within DFT)\cite{kim_spin-orbital_2014}} & $0$ & $-55.7$ & $27.6$ & $7.1$ & $14.5$\\
		\hline
	\end{tabular}
	\caption{The calculated hopping parameters as a function of pressure. For the convention of each parameter, see Fig.~\ref{fig:MLWF} (c). It is found that the lattice optimization does not make significant changes.}
	\label{table_tb}
\end{table*}

\begin{figure*}[!h]
	\includegraphics[scale=0.8]{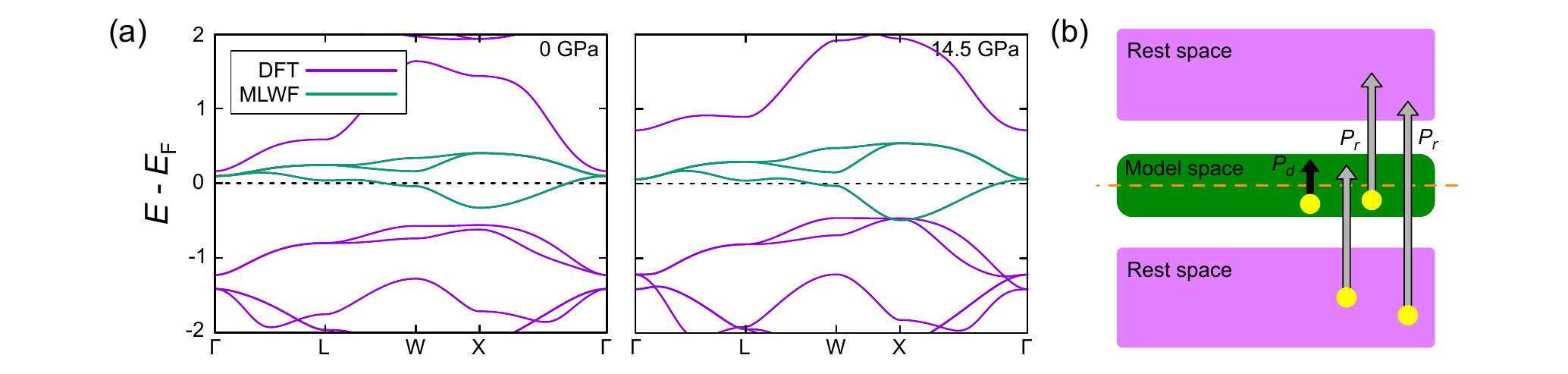}	
	\caption{(a) The calculated band dispersion by DFT-LDA (violet) and MLWF tight-binding parameters (green) at the pressure value of 0 (left) and 14.5 GPa (right). (b) Schematic diagram to describe the polarization in cRPA procedure. The black arrow represents the polarization within the (correlated) model space ($P_d$) while the gray ones refer to the other polarizations ($P_r$). In the current study, the correlated space is defined as molecular $t_2$ orbitals corresponding to the green-colored bands in (a).}
	\label{fig:crpa}
\end{figure*}

\begin{table}[!h]
	\centering
	\begin{tabular}{|l|c|c|c|c|c|c|}
		\hline
		\bf{Pressure (GPa)} & $U$ (eV) & $U'$ (eV) & $J_{\rm{H}}$ (eV) &  $\mathcal{U}$ (eV) & $\bf{\mathcal{U'}}$ (eV) & $\mathcal{J_{\rm{H}}}$ (eV) \\
		\hline
		$\bf{0}$ (ambient) & 0.668 & 0.507 & 0.061 & 0.175 & 0.063 & 0.046 \\ 
		\hline
		$\bf{5}$ & 0.637 & 0.481 & 0.059 & 0.184 & 0.071 & 0.046 \\
		\hline
		$\bf{10}$ & 0.599 & 0.448 & 0.057 & 0.190 & 0.077 & 0.046 \\
		\hline
		$\bf{14.5}$ & 0.585 & 0.436 & 0.056 & 0.193 & 0.080 & 0.046 \\
		\hline
	\end{tabular}
	\caption{The calculated interaction parameters as a function of pressure. $U$, $U'$, and $J_{\rm{H}}$ refers to the intra-, inter-orbital interaction, and the Hund coupling calculated by cRPA, respectively. The corresponding fully screened values are denoted by  $\mathcal{U}$, $\mathcal{U'}$, and $\mathcal{J_{\rm H}}$, respectively.}
	\label{table_cRPA}
\end{table}

In order to consider electron interactions, we performed cRPA calculations which can properly take into account of screening effects in solids \cite{aryasetiawan_frequency-dependent_2004,sasioglu_effective_2011,PhysRevB.77.085122,aryasetiawan_calculations_2006,vaugier_hubbard_2012,sakuma_first-principles_2013,PhysRevB.80.155134,PhysRevB.96.045137} and give rise to the reliable estimation of effective `on-site' interaction strengths being much smaller than the `bare' interactions, $\mathcal{V}$. Within RPA, the fully screened interaction $\mathcal{U}$ can be calculated from
\begin{equation}
\mathcal{U} = \epsilon^{-1}\mathcal{V}
\label{crpa_eq1}
\end{equation}	
where $\epsilon = 1-\mathcal{V}P$ and $P$ is the polarization \cite{aryasetiawan_frequency-dependent_2004}. In order to cooperate with correlated electron models (e.g., Hubbard model), the correlated orbitals or subspaces need to be defined properly. The effective Coulomb interaction $U$ in such a model can be represented by \cite{aryasetiawan_frequency-dependent_2004}
\begin{equation}
U = [1-\mathcal{V}(P-P_d)]^{-1}\mathcal{V} = [1-\mathcal{V}(P_r)]^{-1}\mathcal{V} 
\label{crpa_eq2}
\end{equation}
where $P_d$ and $P_r$ refers to the polarization within the correlated orbitals and the other (`rest') space, respectively; see Fig.~\ref{fig:crpa}. The relationship between the fully screened interaction, $\mathcal{U}$, and the partially-screened (`constrained') $U$ can be found by \cite{aryasetiawan_frequency-dependent_2004}
\begin{align}
\mathcal{U} &= [1-\mathcal{V}(P_r+P_d)]^{-1}\mathcal{V} = [(1-\mathcal{V}P_r)\{1-(1-\mathcal{V}P_r)^{-1}\mathcal{V}P_d\}]^{-1}\mathcal{V} \\
&=\{1-(1-\mathcal{V}P_r)^{-1}\mathcal{V}P_d\}^{-1}(1-\mathcal{V}P_r)^-1\mathcal{V} = [1-UP_d]^{-1}U.
\label{crpa_eq3}
\end{align}	
Our calculation results of these values are presented in Table.~\ref{table_cRPA}.

It is noted that the strengths of `on-site' Coulomb and Hund\textsc{\char13}s interaction are smaller than the typical values for 5$d$ transition metal ions. It is is reasonably well understood from the nature of molecular orbitals which are distributed over the four atomic Ta sites. According to a recent study, these interaction parameters get reduced by a factor of $\sim$1/4 \cite{kim2018molecular}. The effect beyond this simple estimation such as the screenings of other molecular orbitals have been taken into account by our cRPA calculation.

\section{Tight-binding model description}
\label{TB_hopping}

\begin{figure*}[!h]
	\includegraphics[scale=0.4]{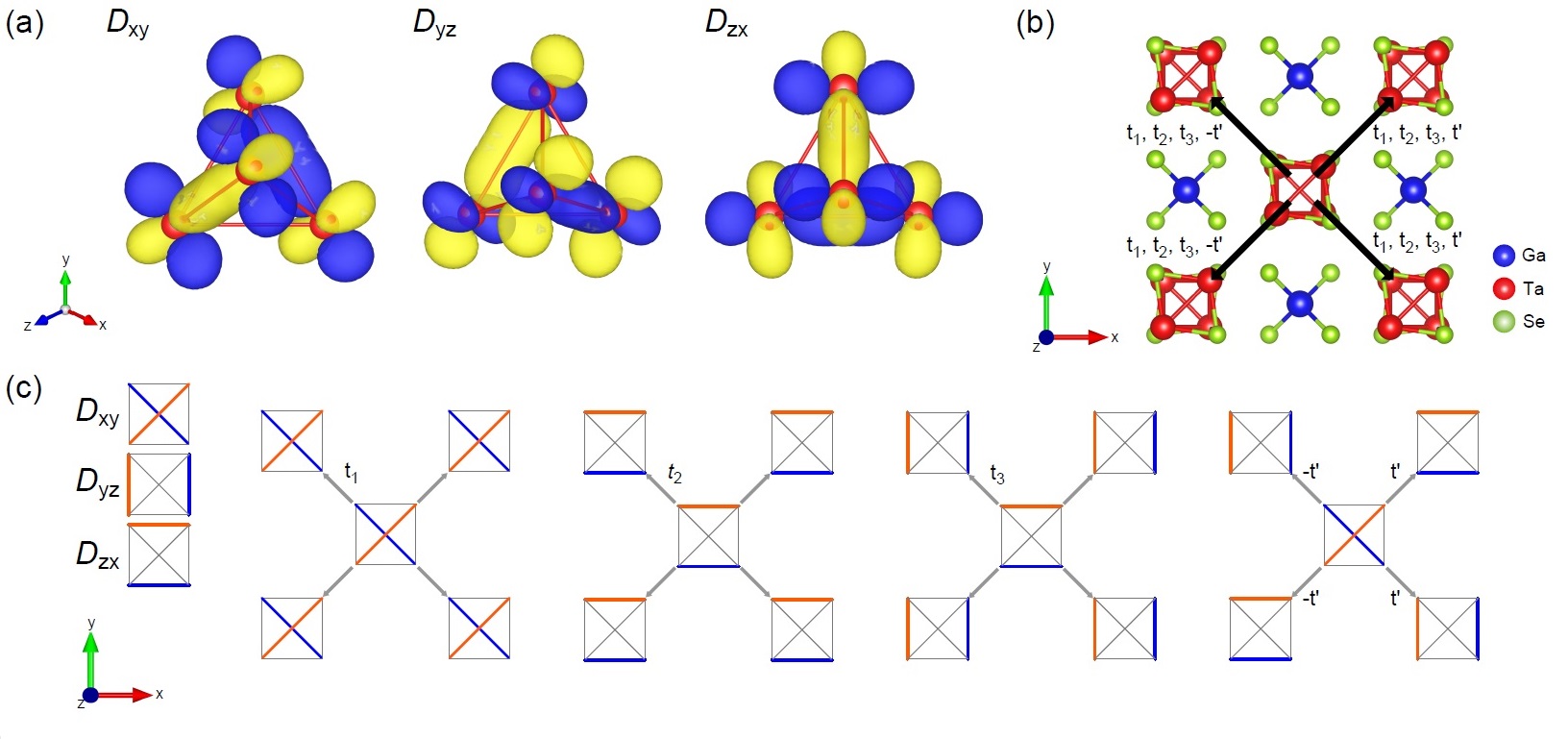}	
	\caption{(a) The calculated MLWFs of three molecular $t_2$ orbitals where we used the isosurface value of 0.05. (b) Top view of crystal structure which clearly shows the arrangement of Ta$_4$Se$_4$ and GaSe$_4$ clusters in the xy-plane. The hoppings in between Ta$_4$Se$_4$ cluster sites are depicted by black arrows which are orbital dependent. (c) The schematic figure to represent the orbital-dependent hoppings. The calculated four major hoppings, t$_1$, t$_2$, t$_3$, and t$'$ are presented in Table.~\ref{table_tb}.}
	\label{fig:MLWF}
\end{figure*}

In this section, we construct the tight binding model description for the completeness. We mainly repeat the description of Ref. \onlinecite{kim_spin-orbital_2014} here. The starting point of our tight-binding model is the molecular $t_2$ orbitals ($D_{xy},D_{yz},D_{zx}$) basis. In the absence of the SOC, the nearest-neighbor hopping matrix from $i$ site to $j$ site can be generally written as,

\begin{equation}
\hat{T}_{{\rm hopping};ij} =
\left(
\begin{array}{ccc}
S_{11} & S_{12} - A_{12} & S_{13} + A_{13}   \\
S_{12} + A_{12} & S_{22} & S_{23} - A_{23}   \\
S_{13}- A_{13} & S_{23} + A_{23} & S_{33}
\end{array}
\right)
\label{eq:H12}
\end{equation}

Here, we separate the inversion even and odd hopping components as $S$ and $A$ respectively. According to the Wannier function analysis, there exist four distinct hopping channels  $t_1$, $t_2$, $t_3$, and $t'$. The inversion even hopping terms, $t_1$, $t_2$, and $t_3$, correspond to $t_{dd1}$ ($\sigma$-type),
$t_{pd}$ ($\pi$-type), and $t_{dd2}$ ($\delta$-type) hopping integrals (See Figure~\ref{fig:MLWF}). In addition,
 the inversion odd term $t'$ is allowed due to the lack of inversion symmetry. In terms of the hopping matrix defined in Eq. \eqref{eq:H12}, the matrix elements are explicitly given as,

\bea
(n_1,n_2,n_3)=({ \pm 1,0,0})
~~ S_{11}=t_1, ~ S_{22}=S_{33}=t_2, ~ S_{23}=-t_3, ~ A_{13}=-A_{12} = \mp t'
\\
\nonumber
(n_1,n_2,n_3)=({ 0, \pm 1,0})
~~S_{11}=S_{33}=t_2, ~ S_{22}=t_1, ~ S_{13}=-t_3, ~ A_{12}=-A_{23} = \mp t'
\\
\nonumber
(n_1,n_2,n_3)=({ 0,0, \pm 1})
~~S_{11}=S_{22}=t_2, ~ S_{33}=t_1, ~ S_{12}=-t_3, ~ A_{13}= -A_{23} = \pm t'
\\
\nonumber
(n_1,n_2,n_3)=({ \pm 1,\mp 1,0})
~~S_{11}=S_{22}=t_2, ~ S_{33}=t_1, ~ S_{12}=t_3, ~ A_{13}=A_{23} = \pm t'
\\
\nonumber
(n_1,n_2,n_3)=({ 0, \pm 1,\mp 1})
~~S_{11}=t_1, ~ S_{22}=S_{33}=t_2, ~ S_{23}=t_3, ~ A_{13}=A_{12} = \pm t'
\\
\nonumber
(n_1,n_2,n_3)=({ \pm 1,0,\mp 1})
~~S_{11}=S_{33}=t_2, ~ S_{22}=t_1, ~ S_{13}=t_3, ~ A_{12}=A_{23} = \mp t'
\eea
where $(n_1,n_2,n_3)$ characterizes the direction of the hopping, ${\bf r}_{ij} = n_1 {\bf a}_1 + n_2 {\bf a}_2 + n_1 {\bf a}_2$. $\bf a_{1,2,3}$ are the unit vectors of the FCC lattice. We now include the SOC effect in the Hamiltonian as,
\begin{equation}
\hat{H}_{\rm SOC}=\lambda_{\rm SO} {\bf L} \cdot {\bf S}.
\label{eq:Hsoc}
\end{equation}
where $\lambda_{\rm SO}$ is the strength of the SOC.
and ${\bf L}$ and ${\bf S}$ are the orbital and the spin angular momentum operators, respectively. We can rewrite the Hamiltonian in Eq. \eqref{eq:H12} in terms of $j_{\textrm{eff}}$ basis:
\begin{eqnarray}
&|&j_{\rm eff}=\frac{1}{2},m=\frac{1}{2} \rangle
= 
\frac{1}{\sqrt{3}}|D_{yz}\downarrow\rangle+\frac{i}{\sqrt{3}}|D_{xz}\downarrow\rangle+\frac{1}{\sqrt{3}}|D_{xy}\uparrow\rangle,
\nonumber
\\
&|& j_{\rm eff}=\frac{1}{2},m=-\frac{1}{2} \rangle
= 
\frac{1}{\sqrt{3}}|D_{yz}\downarrow\rangle-\frac{i}{\sqrt{3}}|D_{xz}\downarrow\rangle-\frac{1}{\sqrt{3}}|D_{xy}\uparrow\rangle,
\nonumber
\\
&|& j_{\rm eff}=\frac{3}{2},m=\frac{3}{2} \rangle
=
-\frac{1}{\sqrt{2}}|D_{yz}\uparrow\rangle-\frac{i}{\sqrt{2}}|D_{xz}\uparrow\rangle,
\nonumber
\\
&|& j_{\rm eff}=\frac{3}{2},m=\frac{1}{2} \rangle
= 
-\frac{1}{\sqrt{6}}|D_{yz}\downarrow\rangle-\frac{i}{\sqrt{6}}|D_{xz}\downarrow\rangle+\sqrt{\frac{2}{3}}|D_{xy}\uparrow\rangle,
\nonumber
\\
&|& j_{\rm eff}=\frac{3}{2},m=-\frac{1}{2} \rangle
=
\frac{1}{\sqrt{6}}|D_{yz}\uparrow\rangle-\frac{i}{\sqrt{6}}|D_{xz}\uparrow\rangle+\sqrt{\frac{2}{3}}|D_{xy}\downarrow\rangle,
\nonumber
\\
&|& j_{\rm eff}=\frac{3}{2},m=-\frac{3}{2} \rangle
= 
\frac{1}{\sqrt{2}}|D_{yz}\downarrow\rangle-\frac{i}{\sqrt{2}}|D_{xz}\downarrow\rangle,
\nonumber
\end{eqnarray}
where the arrows indicate the electron spin. In $j_{\rm eff}$ basis, the hopping matrix and the on-site SOC term transforms as,
\begin{equation}
\hat{T}_{{\rm hopping};ij}  = 
\left(
\begin{array}{cc}
T^{1/2}_{ij} & \Theta_{ij} \\
\Theta_{ij}(-A)^\dagger  & T^{3/2}_{ij}  
\end{array}
\right), \;
\hat{H}_{\rm SOC} = \left(
\begin{array}{cc}
+\lambda_{\rm SO}  I_2 & 0 \\
0  & -\frac{1}{2}\lambda_{\rm SO} I_4   
\end{array}
\right).
\end{equation}
where $I_n$ are $n$-dimensional identity matrix. $T^{1/2(3/2)}$ describes the intraband hopping terms of $j_{\rm eff}$ = 1/2 (3/2) bands, and $\Theta$ represents the interband tunnelings. The explicit forms of the hopping matrices follow Eq. \eqref{eq:H12}. If the energy splitting between the $j_{\rm eff}$ = 1/2 and 3/2 bands, $\frac{3}{2}\lambda_{\rm SO}$,
is large compared to the inter-orbital hopping terms $\Theta$, $j_{\rm eff}$ = 1/2 and 3/2 subsectors effectively decouples.

	\section{Projection to $j_{\rm{eff}}=3/2$ basis}
	\label{Appendix:projection}
	
	\begin{figure*}[!h]
		\includegraphics[scale=0.4]{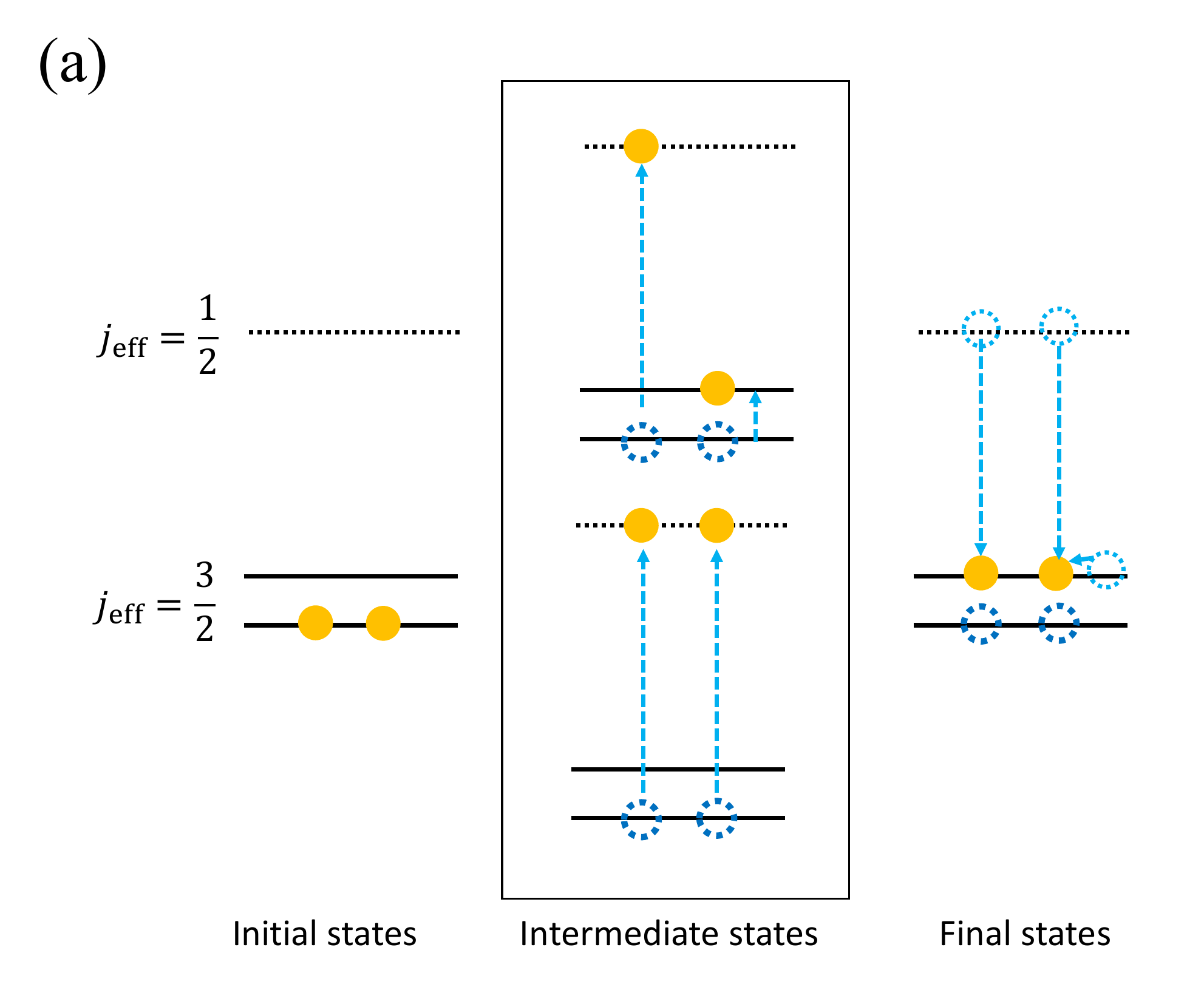}
		\includegraphics[scale=0.7]{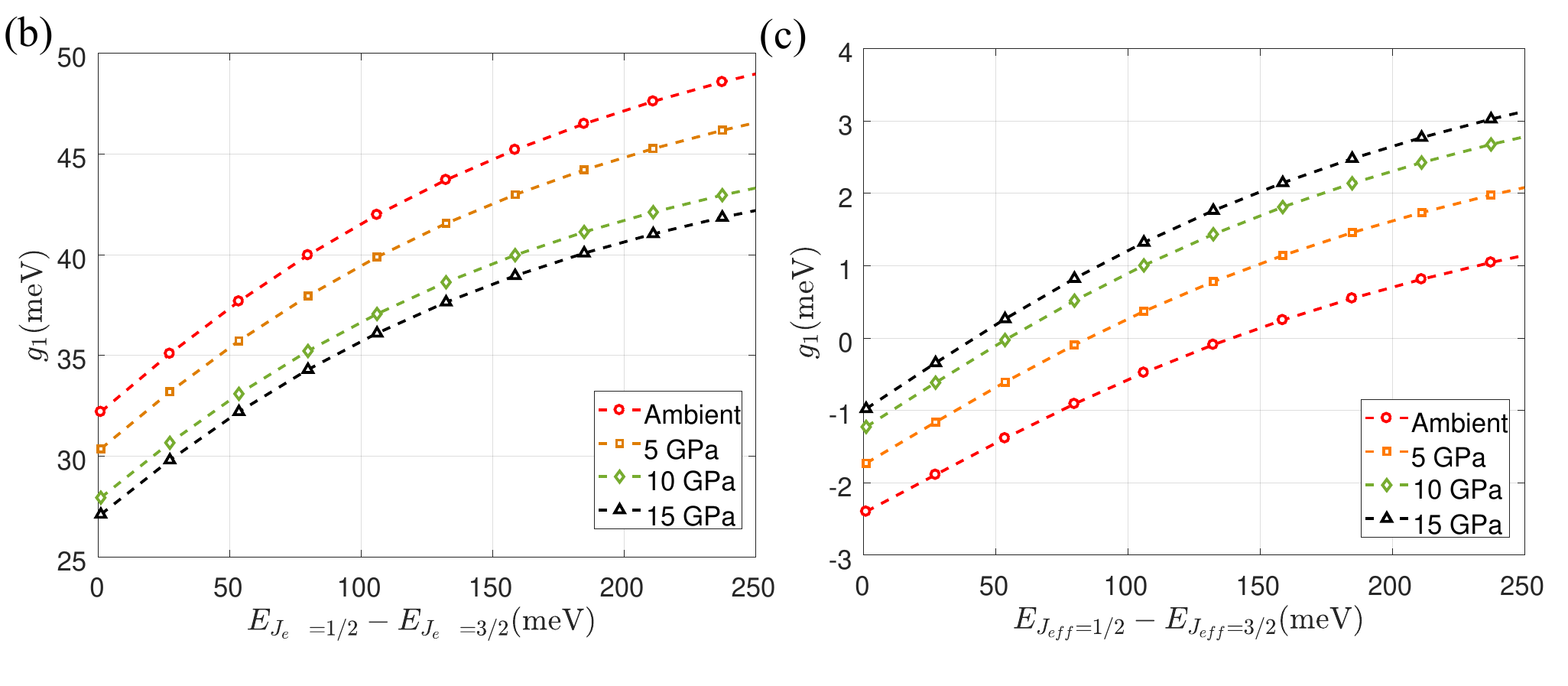}	
		\caption{(a) Schematic representation of the interband tunneling process. The second-order process mediating the intermediate $j_{\rm eff}=1/2$ states contributes to the effective $j_{\rm eff}=3/2$ many-body interactions (b-c) $j_{\rm{eff}}=3/2$ projected pairing interaction strength derived from the exact diagonalization result using (b) the cRPA and (c) the fully screened values of the interaction parameters. We find that the reduction of the band splitting decreases the pairing constants, $g_1$.}
		\label{fig:g1}
	\end{figure*}

	In this section, we project the interacting Hamiltonian to $j_{\rm{eff}}=3/2$ basis. We start our analysis by writing down the interacting Hamiltonian as,
	\begin{align}
	\label{intham}
	H_I&=U\sum_{iu}n_{iu\uparrow}n_{iu\downarrow}+U'\sum_{iu,v<u}n_{iu\sigma}n_{iv\sigma'}\nonumber+\frac{J_H}{2}\sum_{iu\neq v ,\sigma,\sigma'}d^{\dagger}_{iu\sigma}d^{\dagger}_{iv\sigma'}d_{iu\sigma'}d_{iv\sigma}\nonumber
	+\frac{J_H}{2}\sum_{iu\neq v ,\sigma\neq\sigma'}(d^{\dagger}_{iu\sigma}d^{\dagger}_{iu\sigma'}d_{iv\sigma'}d_{iv\sigma}+ h.c.)\nonumber\\
	&=H_U+H_{U'}+H_{J1}+H_{J2}
	\end{align}
	where $i$ is the site index, $u,v\in(xy,yz,zx)$ denote the orbital indices ($D_{xy},D_{yz},D_{zx}$) and $\sigma, \sigma' \in(\uparrow,\downarrow)$ are spin indices. Here, $n_{iu\sigma}=d^\dagger_{iu\sigma}d_{iu\sigma}$ is the number operator and $d_{iu\sigma} ~(d^\dagger_{iu\sigma})$ are the annihilation (creation) operator of electrons at site $i$ and orbital $u$ with spin $\sigma$. 
	$U$ and $U'$ represent the intra-orbital and inter-orbital interaction strengths respectively. The third and fourth terms are the Hund exchange interaction and Hund pair hopping interaction respectively parametrized by $J$. With $(U,U',J)>0$, these interaction terms are repulsive in nature. 
	\subsection{$j_{\rm{eff}}=3/2$ intra-band contribution}
	\label{appendix:intraband}
	Due to spin orbit coupling, in the absence of interactions, the degenerate $t_{2}$ orbital states split into  $J=1/2$ doublet with energy $\lambda$ and $J=3/2$ quartet with energy $-\lambda/2$. Large $\lambda$ leads to the large energy gap between these states with negligible mixing. Hence, we restrict to $J=3/2$ manifold and project the interacting Hamiltonian $H_I$, given in Eq.~\eqref{intham}, to the $J=3/2$ basis states.
	Any operator $O$ expressed in terms of spins and orbitals are projected to the $J=3/2$ subspace through the projection operator $P_{3/2}$ and the projected operator is denoted as $\tilde O\equiv P_{3/2}OP_{3/2}$. Thus, the projection of the number operators $n_{\alpha\beta}$, spin operators ${\bf S}_{\alpha\beta}=(S^{x}_{\alpha\beta}, S^{y}_{\alpha\beta}, S^{z}_{\alpha\beta})$ to the $J=3/2$ subspace can be written as
	\begin{align}
	\label{proj}
	\tilde{n}_{\alpha\beta}&=P_{3/2}n_{\alpha\beta}P_{3/2}=\Psi^\dagger\Big(\frac{3}{4}I-\frac{J^2_\gamma}{3}\Big)\Psi\nonumber\\
	\tilde{S}^\gamma_{\alpha\beta}&=P_{3/2}S^{\gamma}_{\alpha\beta}P_{3/2}=\Psi^\dagger\Big(\frac{3}{4}J_\gamma-\frac{J^3_\gamma}{3}\Big)\Psi\nonumber\\
	\tilde{S}^{\alpha(\beta)}_{\alpha\beta}&=P_{3/2}S^{\alpha(\beta)}_{\alpha\beta}P_{3/2}=\Psi^\dagger\Big(\frac{1}{4}J_{\alpha(\beta)}-\frac{J_\gamma J_{\alpha(\beta)}J_\gamma}{3}\Big)\Psi
	\end{align} 
	where $\alpha,\beta,\gamma\in(x,y,z)$ with $\alpha\neq\beta\neq\gamma$ and the orbital index $u$ in Eq.~\eqref{intham} can be represented as $u=\alpha\beta$. ${\bf J}=(J_x,~J_y,~J_z)$ and $I$ is the $4\times 4$ identity matrix. $\Psi=[d_{3/2}~ d_{1/2}~d_{-1/2}~d_{-3/2}]^T$ is the four component spinor where $d_{m_j}$ is the annihilation operator of electrons in angular momentum state $|J=3/2,m_j\rangle$ and $m_j=(3/2,~1/2,-1/2,-3/2)$ with $J_z$ being diagonal in this basis.
	The projected interacting Hamiltonian is written as,
	\begin{align}
	\label{projintham}
	\tilde H_I&=\tilde H_U+\tilde H_{U'}+\tilde H_{J1}+\tilde H_{J2}
	\end{align}
	which is a $4\times4$ matrix. Hence, we can express the projected interacting Hamiltonian in terms of the  Dirac gamma matrices. In our work, the gamma matrices are explicitly given as following: 
	\bea
	\gamma_{1}=\sigma^z\otimes\sigma^y,~\gamma_{2}=\sigma^z\otimes \sigma^x,~\gamma_{3}=\sigma^y\otimes I_{2\times2},~~\gamma_{4}=\sigma^x\otimes I_{2\times2},~~\gamma_{5}=\sigma^z\otimes \sigma^z
	\eea
	where ${\bf \sigma}=(\sigma^x,~\sigma^y,~\sigma^z)$ are the Pauli matrices. Using the gamma matrices, we can define the time-reversal operator as, $T=\gamma_{1}\gamma_{3}K$ where $K$ is the complex conjugate operator.
	The projected interacting Hamiltonian can, thus, be written as
	\begin{align}
	\label{projham}
	\tilde H_U&=\frac{U}{24}\Big[2(\Psi^\dagger\Psi)^2+(\Psi^\dagger\gamma_4\Psi)^2+(\Psi^\dagger\gamma_5\Psi)^2
	-2(\Psi^\dagger\gamma_{12}\Psi)^2-(\Psi^\dagger\gamma_{34}\Psi)^2-(\Psi^\dagger\gamma_{35}\Psi)^2\Big]\nn\\
	\tilde H_{U'}&=\frac{U'}{12}\Big[4(\Psi^\dagger\Psi)^2-(\Psi^\dagger\gamma_4\Psi)^2-(\Psi^\dagger\gamma_5\Psi)^2\Big]\nn\\
	\tilde H_{J1}&=\frac{-J}{72}\Big[12(\Psi^\dagger\Psi)^2-3(\Psi^\dagger\gamma_4\Psi)^2-3(\Psi^\dagger\gamma_5\Psi)^2-4(\Psi^\dagger\gamma_{12}\Psi)^2-4(\Psi^\dagger\gamma_{13}\Psi)^2
	-(\Psi^\dagger\gamma_{14}\Psi)^2+3(\Psi^\dagger\gamma_{15}\Psi)^2
	-4(\Psi^\dagger\gamma_{23}\Psi)^2\nn\\
	&-(\Psi^\dagger\gamma_{24}\Psi)^2+3(\Psi^\dagger\gamma_{25}\Psi)^2
	+5(\Psi^\dagger\gamma_{34}\Psi)^2-3(\Psi^\dagger\gamma_{35}\Psi)^2\Big]
	-\frac{J}{18}\Big[2(\Psi^\dagger\gamma_{12}\Psi)(\Psi^\dagger\gamma_{34}\Psi)-(\Psi^\dagger\gamma_{23}\Psi)(\Psi^\dagger\gamma_{14}\Psi)\nn\\
	&+(\Psi^\dagger\gamma_{13}\Psi)(\Psi^\dagger\gamma_{24}\Psi)\Big]
	-J\frac{(\Psi^\dagger\gamma_{14}\Psi)(\Psi^\dagger\gamma_{15}\Psi)-(\Psi^\dagger\gamma_{15}\Psi)(\Psi^\dagger\gamma_{23}\Psi)-(\Psi^\dagger\gamma_{13}\Psi)(\Psi^\dagger\gamma_{25}\Psi)-(\Psi^\dagger\gamma_{24}\Psi)(\Psi^\dagger\gamma_{25}\Psi)}{6\sqrt{3}}\nn\\
	\tilde H_{J2}&=\frac{J}{72}\Big[(\Psi^\dagger\Psi)^2+6(\Psi^\dagger\gamma_1\Psi)^2+6(\Psi^\dagger\gamma_2\Psi)^2+3(\Psi^\dagger\gamma_3\Psi)^2+4(\Psi^\dagger\gamma_5\Psi)^2
	-(\Psi^\dagger\gamma_{12}\Psi)^2
	-2(\Psi^\dagger\gamma_{13}\Psi)^2-2(\Psi^\dagger\gamma_{14}\Psi)^2\nn\\
	&-6(\Psi^\dagger\gamma_{15}\Psi)^2-2(\Psi^\dagger\gamma_{23}\Psi)^2-2(\Psi^\dagger\gamma_{24}\Psi)^2
	-6(\Psi^\dagger\gamma_{25}\Psi)^2
	-4(\Psi^\dagger\gamma_{34}\Psi)^2-3(\Psi^\dagger\gamma_{45}\Psi)^2\Big]\nn\\
	&+\frac{J}{18}\Big[(\Psi^\dagger\gamma_{14}\Psi)(\Psi^\dagger\gamma_{23}\Psi)+(\Psi^\dagger\Psi)(\Psi^\dagger\gamma_5\Psi)-(\Psi^\dagger\gamma_{13}\Psi)(\Psi^\dagger\gamma_{24}\Psi)-(\Psi^\dagger\gamma_{12}\Psi)(\Psi^\dagger\gamma_{34}\Psi)\Big]
	\end{align}

	\begin{table}
		\begin{center}
			\begin{tabular}{|l|c|c|c|c|r|}
				\hline
				&$C_4$&$C_3$&$\sigma_h$&$\sigma_v$&$\sigma_d$\\
				\hline
				$\gamma_{1}$&$\gamma_2$&$\gamma_2$&$-\gamma_1$&$-\gamma_1$&$\gamma_2$\\
				\hline
				$\gamma_{2}$&$-\gamma_1$&$\gamma_3$&$-\gamma_2$&$\gamma_2$&$\gamma_1$\\
				\hline
				$\gamma_{3}$&$-\gamma_{3}$&$\gamma_{1}$&$\gamma_3$&$-\gamma_3$&$\gamma_3$\\
				\hline
				$\gamma_{4}$&$-\gamma_4$&$-\frac{\gamma_4+\sqrt{3}\gamma_5}{2}$&$\gamma_4$&$\gamma_4$&$-\gamma_4$\\
				\hline
				$\gamma_{5}$&$\gamma_5$&$\frac{\sqrt{3}\gamma_4-\gamma_5}{2}$&$\gamma_5$&$\gamma_5$&$\gamma_5$\\
				\hline
			\end{tabular}
			\caption{The table shows the transformation of $\gamma_i,~i\in[1,5]$ under $C_4$ and $C_3$ rotations and reflections about the horizontal plane ($\sigma_h$), one of the vertical plane ($\sigma_v$) and dihedral plane ($\sigma_d$).}
			\label{tab4}
		\end{center}
	\end{table}
	
	\subsubsection{Fierz transformation}\label{Appendix:Fierz}
	
	Using Fierz identity we can decompose the particle-hole channel interactions into the pairing channel interactions. We will show that the repulsive particle hole channel interactions can be written in the form of attractive pairing channel terms through Fierz transformation.\cite{boettcher2018unconventional,sim2019topological}
	The required Fierz identity is,
	\bea
	(\Psi^\dagger M\Psi)(\Psi^\dagger N\Psi)=\frac{1}{16}\textbf{Tr}[M^T\Gamma^AN\Gamma^B](\Psi^\dagger\Gamma^A\Psi^*)(\Psi^T\Gamma^B\Psi).
	\label{eq:fierz}
	\eea
	Eq.~\eqref{eq:fierz} are non-zero only for antisymmetric $\Gamma$ matrices, i.e. $\Gamma\in{\gamma_{13},i\gamma_{13}\gamma_1,\gamma_{13}\gamma_2,i\gamma_{13}\gamma_3,\gamma_{13}\gamma_4,\gamma_{13}\gamma_5}$. 
	
	Using these pieces of information, we construct a Table \ref{tab1} giving the values of $4C^{AA}_{N^TN}$ with rows representing the matrix $N$ and column representing the matrix $\Gamma^A$ where $\textbf{Tr}[M^T\Gamma^AN\Gamma^B]=C^{AB}_{M^TN}$.  After the Fierz transformation, the interactions with $M\neq N$ vanishes.
	\begin{table*}
		\begin{center}
			\begin{tabular}{|l|c|c|c|c|c|c|c|c|c|c|c|c|c|c|c|r|}
				\hline
				&$I$&$\gamma_1$&$\gamma_2$&$\gamma_3$&$\gamma_4$&$\gamma_5$&$\gamma_{12}$&$\gamma_{13}$&$\gamma_{14}$& $\gamma_{15}$&$\gamma_{23}$& $\gamma_{24}$& $\gamma_{25}$& $\gamma_{34}$& $\gamma_{35}$& $\gamma_{45}$\\
				\hline
				$\gamma_{13}$&$1$&$1$&$1$&$1$&$1$&$1$&$-1$&$-1$&$-1$& $-1$&$-1$& $-1$& $-1$& $-1$& $-1$& $-1$\\
				\hline
				$i\gamma_{13}\gamma_1$&$1$&$1$&$-1$&$-1$&$-1$&$-1$&$1$&$1$&$1$& $1$&$-1$& $-1$& $-1$& $-1$& $-1$& $-1$\\
				\hline
				$\gamma_{13}\gamma_2$&$1$&$-1$&$1$&$-1$&$-1$&$-1$&$1$&$-1$&$-1$& $-1$&$1$& $1$& $1$& $-1$& $-1$& $-1$\\
				\hline
				$i\gamma_{13}\gamma_3$&$1$&$-1$&$-1$&$1$&$-1$&$-1$&$-1$&$1$&$-1$& $-1$&$1$& $-1$& $-1$& $1$& $1$& $-1$\\
				\hline
				$\gamma_{13}\gamma_4$&$1$&$-1$&$-1$&$-1$&$1$&$-1$&$-1$&$-1$&$1$& $-1$&$-1$& $1$& $-1$& $1$& $-1$& $1$\\
				\hline
				$\gamma_{13}\gamma_5$&$1$&$-1$&$-1$&$-1$&$-1$&$1$&$-1$&$-1$&$-1$& $1$&$-1$& $-1$& $1$& $-1$& $-1$& $1$\\
				\hline
			\end{tabular}
			\caption{The table gives the values of $4C^{AA}_{N^TN}$ with rows representing the matrix $N$ and column representing the matrix $\Gamma^A$.}
			\label{tab1}
		\end{center}
	\end{table*}
	From Table \ref{tab1}, we can write the projected interaction Hamiltonian terms in \eqref{projham} in pairing channel form as,
	\begin{align}
	\tilde H_U&=\frac{U}{24}[2(\Psi^\dagger(\gamma_{13})^\dagger\Psi^*)(\Psi^T\gamma_{13}\Psi)
	+(\Psi^\dagger(\gamma_{13}\gamma_4)^\dagger\Psi^*)(\Psi^T\gamma_{13}\gamma_4\Psi)
	+(\Psi^\dagger(\gamma_{13}\gamma_5)^\dagger\Psi^*)(\Psi^T\gamma_{13}\gamma_5\Psi)]\nn\\
	\tilde H_{U'}&=\frac{U'}{24}[(\Psi^\dagger(\gamma_{13})^\dagger\Psi^*)(\Psi^T\gamma_{13}\Psi)
	+3(\Psi^\dagger (\gamma_{13}\gamma_1)^\dagger\Psi^*)(\Psi^T\gamma_{13}\gamma_1\Psi)
	+3(\Psi^\dagger(\gamma_{13}\gamma_2)^\dagger\Psi^*)(\Psi^T\gamma_{13}\gamma_2 \Psi)\nn\\
	&+3(\Psi^\dagger (\gamma_{13}\gamma_3)^\dagger\Psi^*)(\Psi^T\gamma_{13}\gamma_3\Psi)
	+2(\Psi^\dagger(\gamma_{13}\gamma_4)^\dagger\Psi^*)(\Psi^T\gamma_{13}\gamma_4\Psi)
	+2(\Psi^\dagger(\gamma_{13}\gamma_5)^\dagger\Psi^*)(\Psi^T\gamma_{13}\gamma_5\Psi)]\nn\\
	\tilde H_{J1}&=\frac{-J}{72}[3(\Psi^\dagger (\gamma_{13})^\dagger\Psi^*)(\Psi^T\gamma_{13}\Psi)
	+3(\Psi^\dagger (\gamma_{13}\gamma_1)^\dagger\Psi^*)(\Psi^T\gamma_{13}\gamma_1\Psi)
	+3(\Psi^\dagger (\gamma_{13}\gamma_2)^\dagger\Psi^*)(\Psi^T \gamma_{13}\gamma_2\Psi)\nn\\
	&+3(\Psi^\dagger (\gamma_{13}\gamma_3)^\dagger\Psi^*)(\Psi^T\gamma_{13}\gamma_3\Psi)
	+6(\Psi^\dagger (\gamma_{13}\gamma_4)^\dagger\Psi^*)(\Psi^T\gamma_{13}\gamma_4\Psi)
	+6(\Psi^\dagger (\gamma_{13}\gamma_5)^\dagger\Psi^*)(\Psi^T\gamma_{13}\gamma_5\Psi)]\nn\\
	\tilde H_{J2}&=\frac{J}{72}[12(\Psi^\dagger(\gamma_{13})^\dagger\Psi^*)(\Psi^T\gamma_{13}\Psi)
	-3(\Psi^\dagger(\gamma_{13}\gamma_4)^\dagger\Psi^*)(\Psi^T\gamma_{13}\gamma_4\Psi)
	-3(\Psi^\dagger(\gamma_{13}\gamma_5)^\dagger\Psi^*)(\Psi^T\gamma_{13}\gamma_5\Psi)]
	\label{fierzham}
	\end{align}
	The total projected pairing interaction can be written as
	\begin{align}
	\tilde H_{I}&=\frac{2U+U'+3J}{24}(\Psi^\dagger\gamma_{13}^\dagger\Psi^*)(\Psi^T\gamma_{13}\Psi)+
	\frac{3U'-J}{24}[(\Psi^\dagger(\gamma_{13}\gamma_1)^\dagger\Psi^*)(\Psi^T\gamma_{13}\gamma_1\Psi)+(\Psi^\dagger(\gamma_{13}\gamma_2)^\dagger\Psi^*)(\Psi^T\gamma_{13}\gamma_2\Psi)\nonumber\\
	&+(\Psi^\dagger(\gamma_{13}\gamma_3)^\dagger\Psi^*)(\Psi^T\gamma_{13}\gamma_3\Psi)]
	+\frac{U+2U'-3J}{24}[(\Psi^\dagger(\gamma_{13}\gamma_4)^\dagger\Psi^*)(\Psi^T\gamma_{13}\gamma_4\Psi)
	+(\Psi^\dagger(\gamma_{13}\gamma_5)^\dagger\Psi^*)(\Psi^T\gamma_{13}\gamma_5\Psi)]
	\label{totintham}
	\end{align}
	The first term in Eq.~\eqref{totintham} give rise to the singlet pairing while the remaining terms here correspond to the quintet pairing channel. We notice that the quintet pairing channel can be attractive in the strong Hund's coupling limit, whereas the single pairing channel is always repulsive.
	
	\subsection{Effect of Interband coupling between $j_{\rm{eff}}=3/2$ and $j_{\rm{eff}}=1/2$ bands}
	\label{appendix:interband}
	We now derive the effective many-body Hamiltonian induced by the interband coupling between $j_{\rm{eff}}=3/2$ and $j_{\rm{eff}}=1/2$ bands. To systematically calculate the effective Hamiltonian of $j_{\rm{eff}}=3/2$ band, we employ the many-body Schrieffer-Wolff transformation\cite{BRAVYI20112793} and exact diagonalization technique.
	
	\subsubsection{Schrieffer-Wolff transformation}
	
	The Schrieffer-Wolff transformation decomposes the many-body interaction, $H_{I}$, into diagonal and off-diagonal component, $D(H_I)$ and $O(H_I)$ respectively, where each of them acts with the projection operator, $P$, to a target subsystem and its complementary space, $Q=1-P$. The diagonal and off-diagonal part can be written as,
	\bea
	D(H_I)=P H_I P + QH_I Q
	\\
	\nonumber
	O(H_I)=P H_I Q + QH_I P
	\eea
	In our case, we aim to derive the effective many-body interacting Hamiltonian of $j_{\rm{eff}}=3/2$ bands. Therefore, $P$ projects to the target subspace, characterized by the quarter-filled electrons in $j_{\rm{eff}}=3/2$ bands and zero electrons in $j_{\rm{eff}}=1/2$ bands. The complementary space corresponds to the states with $N$ electrons filled in $j_{\rm{eff}}=1/2$ bands and $N$ electrons removed from quarter filled $j_{\rm{eff}}=3/2$ bands.
	
	The effective Hamiltonian can be perturbatively expanded as\cite{BRAVYI20112793},
	\bea
	H_{eff,1} &=P H_{I} P
	\\
	H_{eff,2} &=\frac{1}{2} P [{S}_1 ,O(H_I) ] p
	\nonumber
	\\
	H_{eff,3} &=\frac{1}{2} P [O(H_I), \mathcal{L} [D(H_I) , S_1]] P
	\nonumber
	\eea
	where we define a superoperator as $\mathcal{L}(X)=\sum_{i,j}\frac{\langle i|O(X)|j\rangle}{E_i-E_j} |i\rangle\langle j|$.
	The first-order term is just equal to the intra-band contributions considered in the previous section. The second-order describes the virtual many-body hopping processes as shown in Fig.\ref*{fig:g1}(a). Other higher-order terms describe the more complicated virtual processes.
	
	More specifically, we rewrite the interacting Hamiltonian in $j_{\rm{eff}}$ basis, which is written as,
	\bea
	H_I=\sum_{i_{1..4},k_{1},k_2,q}V_{i_1i_2i_3i_4} c^\dagger_{i_1}(k_1+q) c^\dagger_{i_2}(k_2-q)
	c_{i_3}(k_2)c_{i_4}(k_1)
	\eea
	where $\vec{i}=(|\frac{1}{2} \frac{1}{2}\rangle,|\frac{1}{2} -\frac{1}{2}\rangle,|\frac{3}{2} \frac{3}{2}\rangle,|\frac{3}{2} \frac{1}{2}\rangle,|\frac{3}{2} -\frac{1}{2}\rangle,|\frac{3}{2} -\frac{3}{2}\rangle)$ indicates the $j_{\rm{eff}}$ basis. The second order effective Hamiltonian can be written as,
	\bea
	H_{eff,2}=\sum_{i_{1..4},k_{1},k_2,q}V_{eff|i_1i_2i_3i_4}(k_1,k_2,q) c^\dagger_{i_1}(k_1+q) c^\dagger_{i_2}(k_2-q)
	c_{i_3}(k_2)c_{i_4}(k_1)
	\eea
	where the coupling constants can be computed from the eigenstates and the energy eigenvalues of the non-interacting Hamiltonian as,
	\bea
	V_{eff,2 | i_1,i_2,i_3,i_4}(k_1,k_2,Q)
	&=2\sum_{j_{1,2,3,4},q_2}
	(\frac{ \theta(\epsilon_{j_3}(k_2-q_2)) \theta(\epsilon_{j_4}(k_1+q_2))}{-\epsilon_{j_3}(k_2-q_2)-\epsilon_{j_4}(k_1+q_2)})
	W_{i_1,i_2,j_3,j_4}(k_1+q_2,k_2-q_2,Q-q_2)
	\\
	&\times 
	W_{j_1,j_2,i_3,i_4}(k_1,k_2,q_2)
	(\delta_{j_3,j_2}\delta_{j_4,j_1})OR(j_3 \leq 2, j_4\leq 2)
	\nonumber
	\eea
	where $W(k_1,k_2,q)(k_1,k_2,q)$ is the interaction coefficient in the energy eigenstate basis. $\theta$ is the heavi-side step function. $\epsilon_i$ is the energy eigenvalue of $i$-th non-interacting band. By numerically  plugging in the information of the non-interacting bands, we find that the second-order interband tunneling always contributes as the attractive singlet and quintet pairing interactions. This result becomes analytically apparent if we consider the flat band limit.  In this limit, $H_{\rm eff,2}$ can be explicitly calculated as,
	\bea
	g_{\rm eff,0}=-\frac{1}{18}\frac{(3J_H+U-U')^2}{\Delta E} \quad g_{\rm eff,1}=-\frac{1}{6}\frac{ J_H^2}{\Delta E}, \quad g_{\rm eff,2}=-\frac{1}{24}\frac{(U-U')^2}{\Delta E},
	\eea
	where $\Delta E$ is the band splitting between $j_{\rm{eff}}=3/2$ and $1/2$ bands. As a result, we find that the leading order corrections are all negative, which contributes as an attractive pairing channel. This result is irrespective of the specific values of $(U,U',J_H)$, since they are the complete square form.
	
	The exact interband tunneling contribution can be numerically computed up to all orders using the exact-diagonalization technique in the flat band limit. Fig. \ref{fig:g1}(b) shows the effective pairing strength derived from the exact diagonalization technique as a function of the interband splitting between $j_{\rm{eff}}=3/2$ and $j_{\rm{eff}}=1/2$. We find that the decrease of the interband spacing increases the effect of the interband tunneling, finally contributing as negative correction of $g_1$ Especially, when the fully screened interaction is taken into account, we find that the pairing interaction strength can turn to a negative value. Eventually, the attractive superconducting pairing channels open. In results, we conclude that the reduction of the band splitting in addition to the strong Hund's coupling may induce the quintet pairing superconductivity in $j_{\rm eff}=3/2$ bands.
	
	\section{mean-field energy calculation}\label{appendix:meanfield}
	
	\begin{figure*}[!h]
		\includegraphics[scale=0.7]{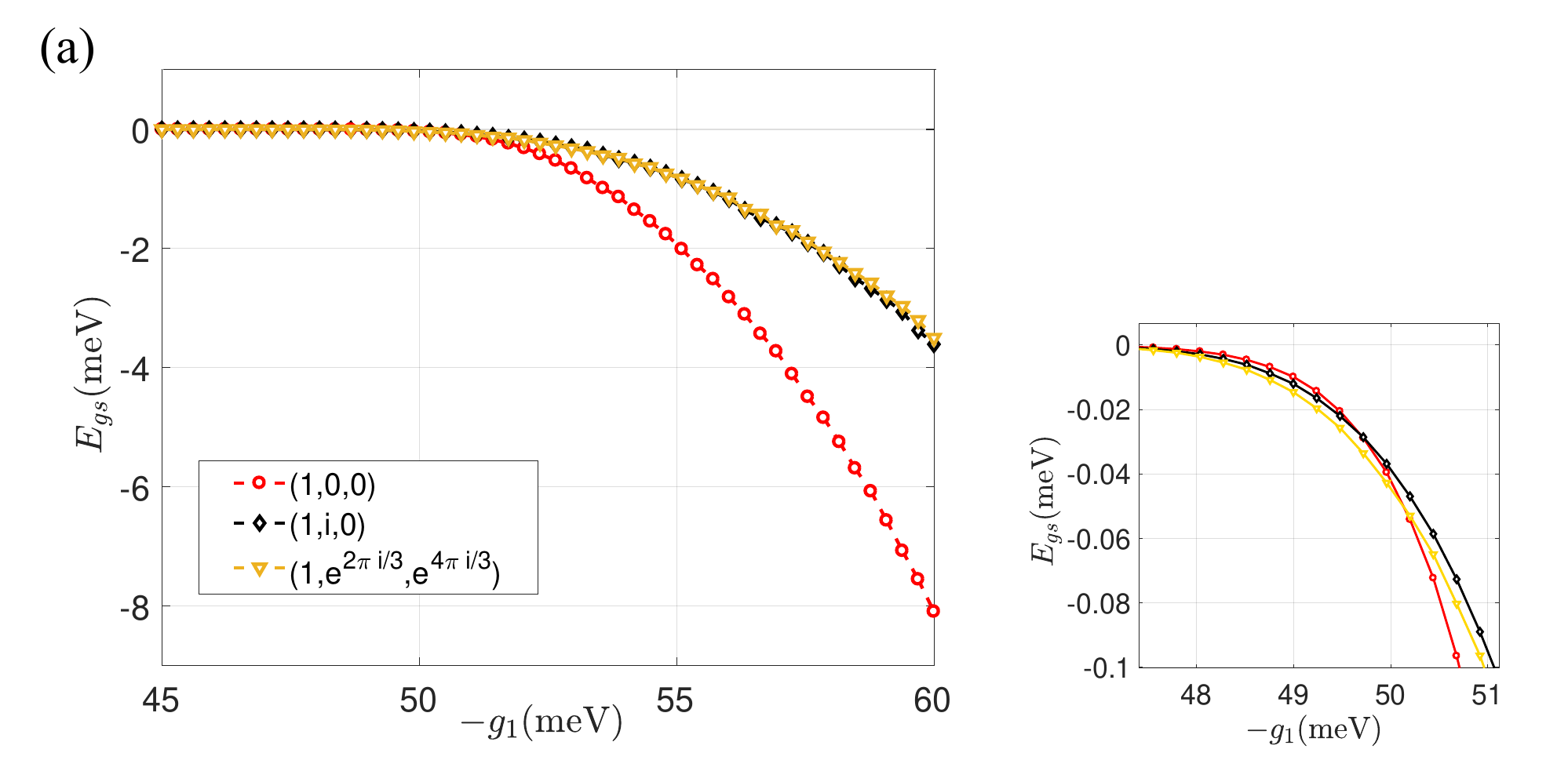}
		\includegraphics[scale=0.5]{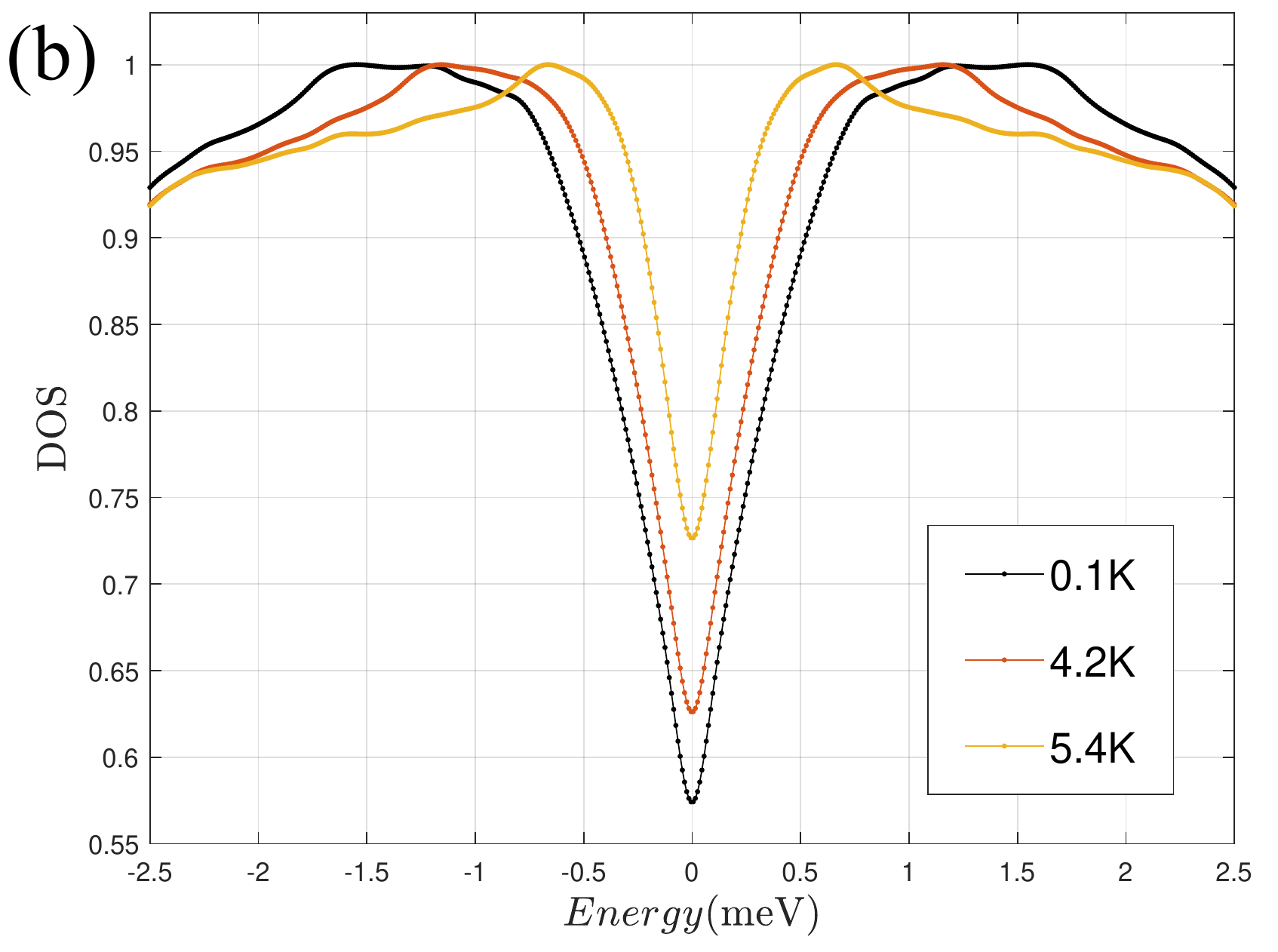}
		\includegraphics[scale=0.4]{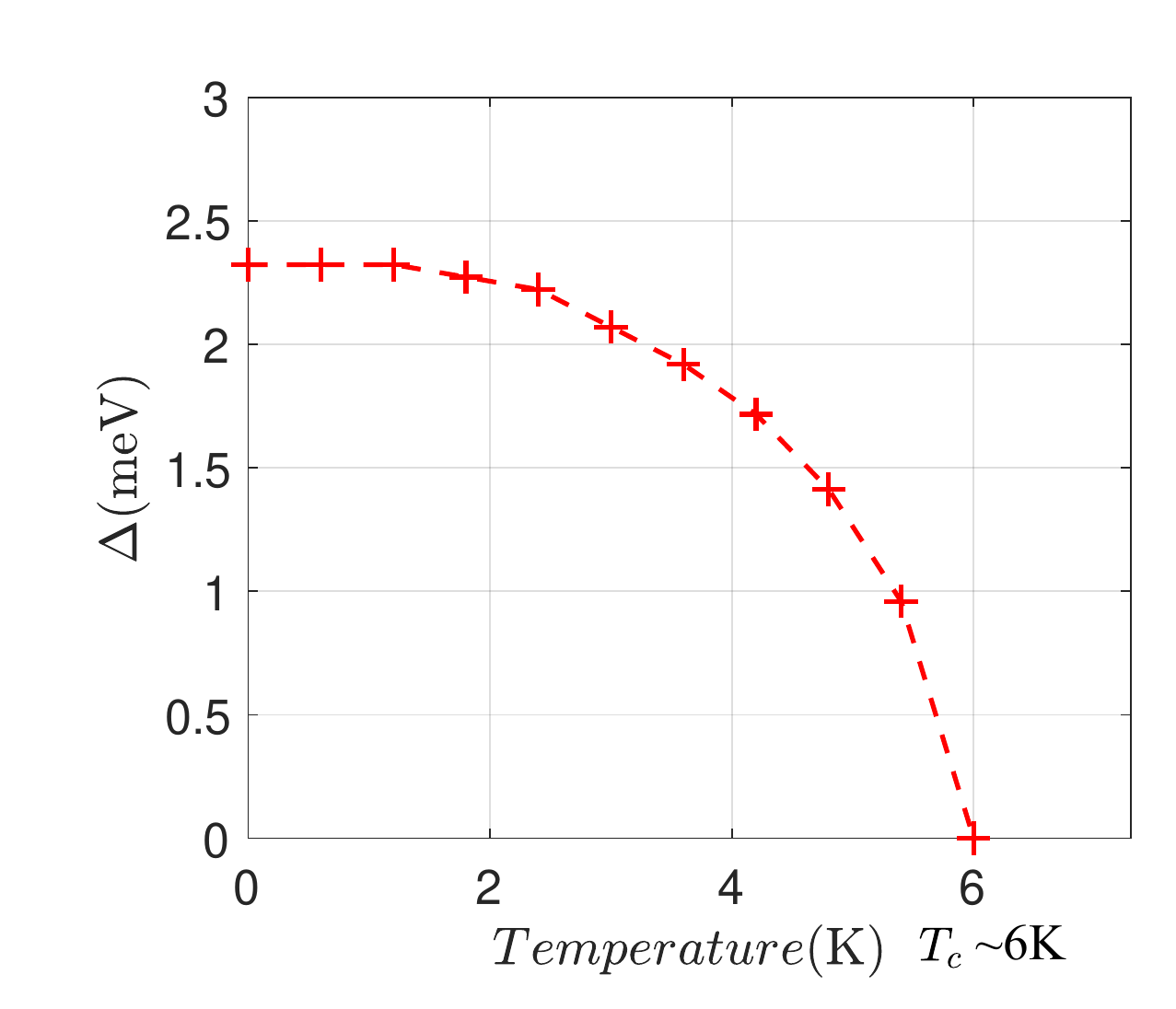}
		\caption{(a) Total energy calculation derived from the mean-field theory as a function of $g_1$. The superconducting order parameters are represented as three component vectors. Inset: magnified figure around $-g_1\approx 50$meV. we find the phase transition from the time-reversal broken phase to the time-reversal symmetric phase. (b) The density of states of $(1,0,0)$-pairing as a function of the temperature. Nodal line superconductivity shows the linear DOS profile. Inset: (1,0,0) order parameter dependence as a function of the temperature. Using the critical temperature $T_c\sim 6$K, we expect the formation of $2.4$meV superconducting order parameter at zero temperature.}
		\label{fig:energetics}
	\end{figure*}
	
	In the previous section, we find that the pairing channel with $t_{2g}$ symmetry can be attractive when the high pressure is applied. Based on this observation, we perform the standard mean-field theory calculation of the superconductivity. We compare the possible order parameter configurations with $t_{2g}$ symmetry: time-reversal symmetric ($1,0,0$)-state, time-reversal broken ($1,i,0$)-state, and ($1,e^{2\pi i /3},e^{4\pi i /3}$)-state. Fig. \ref{fig:energetics}(a) compares the corresponding ground state energies. In the weak coupling limit where $-g_1<50$meV, we find that the time-reversal broken $(1,e^{2\pi i/3},e^{4\pi i/3})$–state is the most energetically favored ground state(See inset for the magnified figure). As $g_1$ is further increased, the phase transition from the time-reversal broken phase to the time-reversal symmetric phase is observed. In particular, GaTa$_4$Se$_8$ has the critical temperature of $T_c\sim 5.8$K, where the corresponding energy scale,  $E_{gs}\sim 0.5$meV, is well outside the phase transition point in Fig. 1 (b). Therefore, we conclude that the time-reversal symmetric $(1,0,0)$-pairing is the relevant ground state and perform further analysis assuming $(1,0,0)$-state in the main text. 
	
	Nevertheless, we also note that the time-reversal broken states can be realizable near the phase transition where the order parameter is suppressed. In such a case, the gap structure of the time-reversal broken states are generally characterized by the Bogoliubov Fermi surfaces. This time-reversal broken states have been similarly found in the previous studies of Luttinger semimetal\cite{PhysRevB.100.224505,PhysRevB.96.214514,PhysRevLett.120.057002,PhysRevLett.118.127001,PhysRevB.99.054505,PhysRevX.8.011029,PhysRevLett.116.177001}. However, we also note the important difference with the Luttinger semimetal that GaTa$_4$Se$_8 $is a quarter filled material where the Fermi level lies at the center of the $j_{\rm eff}=3/2$ valence band.

	As explained in the main text, $(1,0,0)$ pairing state is characterized by $d_{\rm xy}$-wave nodal line gap structure. Fig. \ref{fig:energetics} (b) shows the density of states(DOS) profile of $(1,0,0)$ pairing state with $T_c=5.8$K as a function fo the temperature. Due to the presence of the nodal lines, the DOS profile shows a linear nodal behavior rather than the full gap. This difference with the BCS superconductivity can be directly observed from the tunneling spectroscopy. 
	
	\section{Details of transport calculation}
	
	To model the Josephson junction, we introduce the four band model of the $s$-wave BCS superconductor. The BdG Hamiltonian describing the BCS superconductor is given as,
	\bea
	H_{BCS}(\mathbf{k})=
	\begin{pmatrix}
		[-2t_0(cos(k_x)+cos(k_y)+cos(k_z))-\mu]I_4 & |\Delta|e^{i\phi} \gamma_{13} \\
		|\Delta|e^{-i\phi} \gamma_{13}& -[-2t_0(cos(k_x)+cos(k_y)+cos(k_z))-\mu]I_4
	\end{pmatrix},
	\eea 
	where $\mu$ is the chemical potential. $\phi$ is the order parameter phase difference between the lacunar spinel and the BCS superconductor. To model the tunneling junction, we introduce the tunneling term between the Lacunar spinel and the BCS superconductor as,
	\bea
	H_{tunneling}=\sum_{i\in junction} t_0  \Psi^\dagger_i I_4 c_i,
	\eea 
 	where $\Psi$ and $c$ indicate the four component spinors of the lacunar spinel and the BCS superconductor. the site index $i$ is summed over the junction region. After constructing the tight-binding model of the Josephson junction, we numerically diagonalize the occupied energy of the tight binding model while varying the phase difference, $\phi$. This procedure gives the free energy, $F(\phi)$, as a function of $\phi$, and the Josephson current is given as $ I_J(\phi) =\frac{2e}{\hbar} \frac{d F(\phi)}{d\phi}$. Finally, we derive the current-phase relation of the planar Josephson junction by gradually rotating the orientation of the junction.
	
\end{widetext}

\newpage

\bibliography{biblio2_1}

\end{document}